\documentclass[11pt]{article}   	
\usepackage{geometry}
\usepackage{amsmath}              		
\usepackage{float}
\usepackage{graphicx}
\usepackage{float}
\usepackage{subfigure}
\usepackage{amssymb, amsthm}
\usepackage{calc}

\usepackage{chngcntr}
\counterwithout{figure}{section}
\usepackage{authblk}

\usepackage{lscape}
\usepackage{afterpage}

\usepackage{color,xcolor,ucs}
\usepackage{mathtools}
\usepackage{mathrsfs}
\usepackage{tikz}
\usetikzlibrary{shapes,shadows,arrows}

\title{Estimating the reproductive number, total outbreak size, and reporting rates for Zika epidemics in South and Central America}
\author[$\dag$]{Deborah P. Shutt}
\author[$\ddagger \S$]{Carrie A. Manore}
\author[$\dag$]{Stephen Pankavich}
\author[$\dag$]{Aaron T. Porter}
\author[$\sharp$]{Sara Y. Del Valle}

\affil[$\dag$]{Department of Applied Mathematics and Statistics, Colorado School of Mines, Golden, CO 80401}
\affil[$\ddagger$]{Theoretical Biology and Biophysics, Los Alamos National Laboratory, Los Alamos, NM 87544}
\affil[$\sharp$]{Information Systems and Modeling, Los Alamos National Laboratory, Los Alamos, NM 87544}
\affil[$\S$]{The New Mexico Consortium, Los Alamos, NM 87544 }

\begin{document}

\maketitle

\begin{abstract}
As South and Central American countries prepare for increased birth defects from Zika virus outbreaks and plan for mitigation strategies to minimize ongoing and future outbreaks, understanding important characteristics of Zika outbreaks and how they vary across regions is a challenging and important problem. We developed a mathematical model for the 2015 Zika virus outbreak dynamics in Colombia, El Salvador, and Suriname. We fit the model to publicly available data provided by the Pan American Health Organization,  using Approximate Bayesian Computation to estimate parameter distributions and provide uncertainty quantification. An important model input is the at-risk susceptible population, which can vary with a number of factors including climate, elevation, population density, and socio-economic status. We informed this initial condition using the highest historically reported dengue incidence modified by the probable dengue reporting rates in the chosen countries. The model indicated that a country-level analysis was not appropriate for Colombia. We then estimated the basic reproduction number, or the expected number of new human infections arising from a single infected human, to range between 4 and 6 for El Salvador and Suriname with a median of 4.3 and 5.3, respectively. We estimated the reporting rate to be around 16\% in El Salvador and 18\% in Suriname with estimated total outbreak sizes of 73,395 and 21,647 people, respectively. The uncertainty in parameter estimates highlights a need for research and data collection that will better constrain parameter ranges.

\end{abstract}

\section{INTRODUCTION}
Mosquito-borne diseases contribute significantly to the overall morbidity and mortality caused by infectious diseases in Central and South America. Newly emergent pathogens, such as Zika virus in $2015$, highlight the need for data and models to understand the public health impact of associated diseases and develop mitigation strategies to combat their spread.
In particular, since Zika virus is a newly emergent pathogen in the Americas, its impact on the na{\"i}ve population is relatively unknown.

The disease was first discovered in isolation from a rhesus macaque in the Zika forest of Uganda in 1974 \cite{WHOZika2}. While infrequent human cases were confirmed in later years in both Africa and Southeast Asia, it was not until April 2007 that an outbreak outside of these traditional areas occurred on Yap Island in the North Pacific \cite{duffy2009zika}, and this was followed by another outbreak occurring in French Polynesia beginning October of 2013 \cite{kucharski2016transmission}.  However, the most significant Zika outbreak began within Central and South America in $2015$ \cite{WHOZika} and is currently ongoing. Thus far it has resulted in an estimated 714,636 infections, which includes both suspected and confirmed cases within Latin American and Non-Latin Caribbean countries (accessed Jan 10, 2017) \cite{PAHO}.
This study focuses on the behavior of the current burgeoning epidemic in Colombia, El Salvador, and Suriname.

 Zika is transmitted to humans primarily through bites from infected {\it{Aedes aegypti}} and {\it{Aedes albopictus}} mosquitoes. The transmission is in both directions, that is, infected mosquitoes infect humans and infected humans infect mosquitoes. Upon transmission of the virus from mosquito to human, an individual will become infectious within 3 to 12 days.
Symptoms of infection include fever, rash, joint pain, conjunctivitis, muscle pain and headache.  Recovery from Zika virus disease may require anywhere from 3 to 14 days after becoming infectious, but once contracted humans are immune from the virus for life.  Many people infected with Zika may be asymptomatic 
or will only display mild symptoms that do not require medical attention.  An estimated 80\% of persons infected with Zika virus are asymptomatic \cite{petersen2016interim,duffy2009zika,CDCwebsite2}. Thus, there is a high occurrence of under-reporting for confirmed or suspected Zika cases.  In fact, the number of infected individuals who report their symptoms is estimated to be between $7\%$ and $17\%$ \cite{kucharski2016transmission} of the total number infected by the virus.

Zika can also be transmitted vertically, as a mother can pass the virus to her child during pregnancy, and this can lead to a variety of developmental issues. Most notably, Zika is a cause of microcephaly and other severe fetal brain defects \cite{CDCwebsite}.  Recent evidence further suggests an association between the virus and a higher incidence of Guillain-Barr{\'e} syndrome, a disease in which the immune system damages nerve cells causing muscle weakness and sometimes paralysis \cite{cao2016guillain}.  The potential for sexual transmission of Zika virus has also been confirmed \cite{d2016evidence}.  For instance, the Center for Disease Control (CDC) has determined that Zika can remain in semen longer than in other body fluids, including vaginal fluids, urine, and blood.  It should be noted, however, that while these latter means of transmission exist, the number of new human infections produced in this way is relatively low compared to mosquito-borne infections \cite{cauchemez2016association,towersestimation}, and therefore, in the interest of simplicity and reducing the number of parameters to fit, we will neglect them in formulating a model.

In general, mathematical modeling has been extensively used to understand disease dynamics and the impact of mitigation strategies \cite{hethcote2000mathematics}. A few recent papers have developed models that focus on the behavior of the Zika epidemic.  For example, Gao et al. \cite{gao2016prevention}, proposed an SEIR/SEI model to understand the effects of sexually transmitted Zika in Brazil, Colombia, and El Salvador.  The authors did not, however, distinguish between countries as they
assume that the three nations of interest share common parameter values.  Additionally, Towers et al. \cite{towersestimation} presented a model that incorporates spatial heterogeneity in populations at a granular level.  Their model focused specifically on the spread of Zika within Barranquilla, Colombia and included a sexual transmission term, but concluded that the effects are not significant enough to sustain the disease in the absence of mosquitoes. An approximation of the basic reproductive number was obtained using maximum likelihood methods.  Finally, a study of the impact of short term dispersal on the dynamics of Zika virus is analyzed in \cite{moreno2016role}.  The model formulated within \cite{moreno2016role} does distinguish between asymptomatic and symptomatic infected populations, and focuses on the estimation of the reproductive number between two close communities.

In contrast to previous studies, we adopt a more global approach to understanding the dynamics of the current Zika epidemic and present a model that can be used to study disease transmission at the country level.  We identify differences between countries in regards to parameter values and reproductive numbers.  We discuss the appropriateness of country level analysis for Colombia, El Salvador, and Suriname and quantify the uncertainty within the resulting biological parameter values.

In the following section, we develop a Susceptible-Exposed-Infectious-Recovered (SEIR) /SEI type model which distinguishes between the reported infected population, who are considered symptomatic, and the unreported infected population, who may be asymptomatic or experience symptoms that are not severe enough to seek medical attention.  The data available from the Pan American Health Organization (PAHO) serves to motivate the use of a split infectious population.  The number of reported cases of Zika, both suspected and confirmed, is reported from PAHO by country.  Hence, we develop a model for three countries of interest assuming that the dynamics of the disease within each may be associated with different parameter values. In Section 3, we address the additional complication that the total population of a country cannot be used as the initial susceptible population due to the biology and bionomics of the {\textit{Aedes}} species and human contact with them, varying in regards to temperature, humidity \cite{fay1964biology}, sanitation, demographics and elevation \cite{lozano2012dengue}.  Not everyone within a country is equally susceptible to the Zika infection due to geographic diversity; hence, we calculate the unique at-risk population within each country to use as the initial susceptible population. Using the deterministic model at baseline parameter values, the at-risk population we compute yields realistic initial conditions that compare well with PAHO data. With the initial population sizes for the epidemic in each country known and fixed, we embed the deterministic system into a stochastic process to quantify the uncertainty of the parameter values. This method allows us to use biologically valid parameter ranges and Approximate Bayesian Computation (ABC) methods to obtain parameter distributions that are distinct to each country.  Section 4 provides a detailed explanation of our implementation of the ABC method along with the strengths and weaknesses discovered in applying this method to our particular model.  We summarize the results of the obtained posterior distributions in Section 5. The last section is dedicated to discussion and conclusions.

\section{A Deterministic Zika Model}
\subsection{Vector-borne SEIR Model with Case Reporting}
\tikzstyle{line} = [draw, -latex']
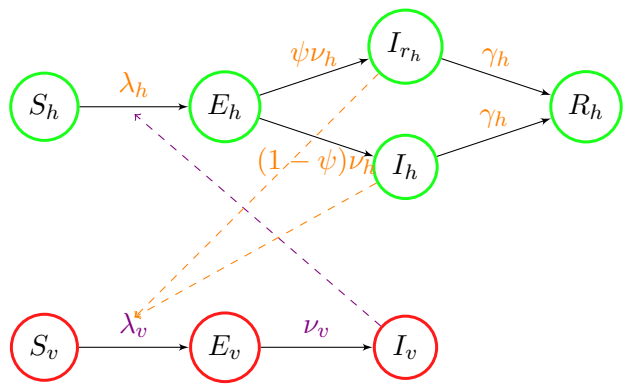
\begin{figure}[H]
\begin{center}
\begin{tikzpicture}[place/.style={circle,draw=red!90,line width=0.4mm, fill=white},
   place2/.style={circle,draw=green!90,line width=0.4mm, fill=white},
   transition/.style={->, line = linetype width=0.4mm},
   trans/.style={->,dashed= linetype width = 0.4mm, violet},
   trans2/.style={->,dashed= linetype width = 0.4mm, orange},scale = 0.4]
\node[place2] (sh) at (-12, 7) {$S_h$};
\node[place2] (eh) at (-6,7) {$E_h$};
\node[place2] (irh) at (0, 9) {$I_{r_h}$};
\node[place2](ih) at (0, 5) {$I_h$};
\node[place2](rh) at (6, 7) {$R_h$};
\node[place](sv) at (-12,-1) {$S_v$};
\node[place](ev) at (-6,-1) {$E_v$};
\node[place](iv) at (0,-1) {$I_v$};

\path
   (sh) edge[transition]node[above,orange]{$\lambda_h$} (eh)
   (eh) edge[transition]node[above,orange]{$\psi  \nu_h$}(irh)
   (eh) edge[transition]node[below,orange]{$(1-\psi ) \nu_h$}(ih)
   (irh) edge[transition]node[above,orange]{$\gamma_h$}(rh)
   (ih) edge[transition]node[above,orange]{$\gamma_h$}(rh)
   (sv) edge[transition]node[above,violet]{$\lambda_v$}(ev)
   (ev) edge[transition]node[above,violet]{$\nu_v$}(iv)
   (iv) edge[trans] (-9,6.75)
   (irh) edge[trans2] (-9,0)
   (ih) edge[trans2] (-9,0)
;

\end{tikzpicture}
\end{center}
\caption{\footnotesize{A schematic representation of \eqref{ode}, modeling the progression of Zika in human populations, denoted by green circles, and mosquito populations, represented with red circles. Susceptible humans start in $S_h$, and move to $E_h$, the exposed population, once infected by a mosquito carrying the virus.  After an intrinsic incubation period, exposed individuals become infectious and transition to either the reported infectious population, $I_{r_h}$, or the unreported infectious population, $I_r$.  Infectious humans will then move to and remain in $R_h$ after recovering from the infection.  The susceptible mosquito population is denoted $S_v$.  After transmission occurs from bitting an infectious human, susceptible mosquitoes transition to the exposed population, $E_v$.  The end of the extrinsic incubation period marks the exposed mosquitoes shift to the infectious class $I_v$, where they remain infectious until death.}}
\label{schematic}
\end{figure}

The spread of Zika relies primarily on interactions between humans and mosquitoes.  In Figure \ref{schematic}, members of the at-risk human population, $S_h$, are bitten by an infectious mosquito and become exposed, or infected  without yet being infectious, at a rate $\lambda_h$.  Humans within the exposed population, $E_h$,  progress from this state to the infectious compartment at a per capita rate, $\nu_h$.  Note that there is a portion of infectious humans who are either asymptomatic or experience less severe symptoms, and therefore go unreported.  The parameter $\psi$ denotes the proportion of humans who seek medical assistance and are reported as either suspected or confirmed Zika patients.  Thus, there are two categories for infectious humans: $I_{r_h}$, the reported infectious population, and $I_h$, the unreported infectious population.  Both infectious populations then recover at the same per capita rate of $\gamma_h$, where $\frac{1}{\gamma_h}$ is the average time spent as infectious.  Upon recovery,  humans acquire lifetime immunity.  Similarly, a member of the susceptible mosquito population, $S_v$,  becomes exposed at a rate $\lambda_v$ when susceptible mosquitoes bite an infectious human, resulting in transmission.  The exposed mosquitoes, $E_v$, transition to the infectious compartment at a per capita rate $\nu_v$, where $\frac{1}{\nu_v}$ is the average extrinsic incubation period.  Once a mosquito is infectious it will remain so for the duration of its lifespan which can range between $8 - 35$ days \cite{andraud2012dynamic,otero2006stochastic,fay1964biology}.  Finally, because of the long duration of infection for mosquitoes relative to their lifespan, demography (i.e., births and deaths) is included within their population dynamics. In total, the system is described by the following eight coupled, nonlinear ordinary differential equations:

\begin{equation}
\label{ode}
\left. \begin{aligned}
	\frac{dS_h}{dt}&=- \lambda_h(t) S_h\\
	\frac{dE_h}{dt}&=  \lambda_h(t) S_h - \nu_h E_h\\
	\frac{dI_{r_h}}{dt} &= \psi \nu_h E_h - \gamma_h I_{r_h}\\
	\frac{dI_h}{dt}&=(1-\psi)\nu_h E_h -\gamma_h I_h\\
	\frac{dR_h}{dt}&=\gamma_h I_{r_h} + \gamma_h I_h\\
	\frac{dS_v}{dt}&=\mu_v N_v- \lambda_v(t) S_v - \mu_v S_v\\
	\frac{dE_v}{dt}&=\lambda_v(t) S_v -\nu_v E_v -\mu_v E_v\\
	\frac{dI_v}{dt}&=\nu_v E_v - \mu_v I_v.
\end{aligned}
\right \}
\end{equation}

\begin{table}[t]
\begin{center}
\begin{tabular}{| c | l |}
\hline $S_h$ & Number of susceptible humans\\
\hline $E_h$ & Number of infected humans who are not yet infectious\\
\hline $I_{r_h}$ & Number of reported infectious humans\\
\hline $I _h$ & Number of unreported infectious humans\\
\hline $R_h$ & Number of recovered humans\\
\hline $S_v$ & Number of susceptible mosquitos\\
\hline $E_v$ & Number of infected mosquitos who are not yet infectious\\
\hline $I_v$ & Number of infectious mosquitos\\
\hline
\end{tabular}
\caption{{\footnotesize{Description of human and mosquito populations state variables}}}
\label{population}
\end{center}
\end{table}

The population state variables are described within Table \ref{population} and parameters are presented in Table \ref{tab:tab2}.
The quantities $N_h$ and $N_v$ represent the total populations of humans and mosquitoes within the model, and remain constant.
We note that the equation for the evolution of $R_h(t)$ decouples from the system, as it is determined merely by computing the remaining population values.

\begin{table}[H]
\begin{center}
\begin{scriptsize}
\begin{tabular}{| c | l | c | }
\hline $\sigma_v$ & Number of times a single mosquito bites a human per unit time. & \\
	                &~~~  The average number of days between bites is $1/\sigma_v$. & (Time$^{-1}$)\\
\hline $\beta_{hv}$ &   Probability of pathogen transmission from an infectious mosquito to a & \\
			&~~~   susceptible human given that a contact between the two occurs.
	                 &(Dimensionless)\\
\hline $\beta_{vh}$ &   Probability of pathogen transmission from an infectious human to a & \\
	                 &~~~  susceptible mosquito given that a contact between the two occurs.
	                 &(Dimensionless)\\
\hline $\nu_h$ &   Per capita rate of progression of humans from the exposed state to & \\
	                 &~~~  the infectious state. The average duration of the latent period is $1/\nu_h$.
	               &(Time$^{-1}$)\\
\hline $\gamma_h$ &   Per capita recovery rate for humans from the exposed state to the & \\
	               &~~~   infectious state. The average duration of the infectious period is $1/\gamma_h$.
	              &(Time$^{-1}$)\\
\hline $\psi$ &   Proportion of the infected human population that is reported.
	              &(Dimensionless)\\
\hline $\nu_v$ &  Per capita rate of progression of mosquitoes from the exposed state to the infectious state. & \\
	              &~~~ The average duration of the extrinsic incubation period is $1/\nu_v$.
	              & (Time$^{-1}$)\\	
\hline $\mu_v$ &   Density-independent death rate for mosquitoes. & \\
		&~~~ The average lifespan of a mosquito is $1/\mu_v$. &(Time$^{-1}$)\\
\hline $\lambda_h$ &   The force of infection from infected mosquito to susceptible human. & \\
	&~~~ This is defined as & \\
	&~~~  (\# of bites per human by mosquitoes per day) & \\
	&~~~ $\times$ (prob. a human is bitten by an infectious mosquito) & \\
	&~~~ $\times$ (prob. transmission occurs $\vert$ a human was bitten by an infected mosquito) & \\
	&~~~ = ($\sigma_v \frac{N_v}{N_h})(\frac{I_v}{N_v})(\beta_{hv})=\frac{\sigma_v \beta_{hv} I_v}{N_h}$.
	&  (Time$^{-1})$\\
\hline $\lambda_v$ &   The force of infection from infected mosquito to susceptible human. & \\
	&~~~ This is defined as & \\
	&~~~ (\# of bites on a human per mosquito per day) & \\
	&~~~ $\times$ (prob. a mosquito bites an infected human) & \\
	&~~~ $\times$ (prob. transmission occurs $\vert$ the mosquito bit an infected person) & \\
	&~~~ = $(\sigma_v )(\frac{I_h+I_{r_h}}{N_h})(\beta_{vh})=\frac{\sigma_v \beta_{vh}( I_h+I_{r_h})}{N_h}$.
	& (Time$^{-1})$\\
	\hline
\end{tabular}
\caption{{\footnotesize{Description of parameters within the Zika model.  Dimensions are provided in parentheses on the right most side of the table.}}}
\label{tab:tab2}
\end{scriptsize}
\end{center}
\end{table}

\subsection{Basic reproductive number}
We define the basic reproductive number, $\mathscr{R}_0$, as the expected number of secondary infections by a single infectious individual over the duration of the infectious period within a fully susceptible population \cite{manore2014comparing}.  As there is more than one class of infectives involved, we utilize the next generation method to derive an explicit formula for $\mathscr{R}_0$, defined mathematically by the spectral radius of the next generation matrix \cite{diekmann1990definition}. We follow the process in \cite{van2002reproduction} and define $\bold{x} = [E_h, I_{r_h},I_h, E_v, I_v, S_h, R_h, S_v]^T $; thus reordering the presentation of populations from the original system to ensure our calculations possess the correct biological representation.  Let $\mathscr{F}_i(x)$ be the rate of appearance of new infections in compartment \textit{i}.  We indicate the rate of transfer of individuals out of compartment \textit{i} as ${\mathscr{V}}^-_i(x)$  and  the rate of transfer of individuals into compartment \textit{i} by all other means as $\mathscr{V}^+_i(x)$.  Thus, our system can be expressed in a condensed version as $\dot{\bold{x}}_i = \mathscr{F}_i(x) - \mathscr{V}_i(x)$ where $\mathscr{V}_i(x) = {\mathscr{V}}^-_i(x)-\mathscr{V}^+_i(x)$ for $i = 1,..., 8$.\\


Next, we compute $\bold{F} = [\frac{\partial\mathscr{F}_i}{\partial x_j}(\bold{x}_0)]$ and $\bold{V} = [\frac{\partial\mathscr{V}_i}{\partial x_j}(\bold{x}_0)]$ for the exposed and infected compartments, namely for  $1 \leq i,j \leq 5$, where $\bold{x}_0=[0,0,0,0,0,H_0,0,K_v]$ is the disease free equilibrium state with $H_0$ and $K_v$ being the initial population sizes of humans and mosquitoes respectively, and obtain the following $5 \times 5$ matrices:\\
\begin{equation*}
\begin{small}
\centering
\mathbf{F} = \left(
\def\arraystretch{2}\begin{array}{ccccc}
0 & 0 & 0 & 0 & \beta_{hv} \sigma_v\\
0 & 0 & 0 & 0 & 0\\
0 & 0 & 0 & 0 & 0\\
0 & \frac{\beta_{vh} \sigma_v K_v}{H_0} &\frac{\beta_{vh} \sigma_v K_v}{H_0} & 0 & 0\\
0 & 0 & 0 & 0 & 0
\end{array}\right)
\qquad \qquad
\mathbf{V} = \left(
\def\arraystretch{2}\begin{array}{ccccc}
\nu_h & 0 & 0 & 0 & 0\\
-\nu_h \psi & \gamma_h & 0 & 0 & 0\\
-\nu_h(1- \psi)  & 0 &  \gamma_h & 0 & 0\\
0 & 0 & 0 & \mu_v +\nu_v & 0\\
0 & 0 & 0 & -\nu_v & \mu_v
\end{array}\right)
\qquad
\end{small}
\end{equation*}
\\
Hence, we calculate the Reproductive Number as:
$$\mathscr{R}_0 := \rho(FV^{-1}) = \frac{\sigma_v \sqrt{K_v \beta_{hv} \beta_{vh} \nu_v}}{\sqrt{H_0 \gamma_h \mu_v(\mu_v+\nu_v)}}=\sqrt{R_{hv} R_{vh}}$$\\
where $\rho(A)$ represents the spectral radius of the matrix $A$, and we have defined the quantities $R_{hv} = \left(\frac{\nu_v}{\mu_v + \nu_v}\right)\left(\frac{\sigma_v}{\mu_v}\right)\beta_{hv}$ and $R_{vh} = \left(\frac{K_v}{H_0}\right)\left(\frac{\sigma_v}{\gamma_h}\right)\beta_{vh}$.\\

Here, $R_{hv}$ is the expected number of secondary infections in a fully susceptible human population resulting from one newly introduced infected mosquito.  It is composed of the product of three terms.  The first term, $\frac{\nu_v}{\mu_v + \nu_v}$, represents the probability that an exposed mosquito will survive the extrinsic incubation period.  The second term, $\frac{\sigma_v}{\mu_v}$, is the number of human bites an infectious mosquito would make if humans were freely available.  The third term, $\beta_{hv}$, is the probability of transmission occurrence given that a human is bitten by an infected mosquito.  The number of secondary infections in a fully susceptible population of mosquitos resulting from one newly introduced infected human is represented by $R_{vh}$.  This value is also formed by the product of three terms.  The first, $\frac{K_v}{H_0}$, is the vector to host ratio.  The $\frac{\sigma_v}{\gamma_h}$ term is the maximum number of bites an infectious human will experience before recovery without impediment to mosquito bites.  Finally, given that a susceptible mosquito bites an infectious human, $\beta_{vh}$ is the probability of transmission from human to mosquito.  The type reproductive number, or expected number of secondary human cases resulting from one newly infectious human, is ${\mathscr{R}_0}^T := (\mathscr{R}_0)^2$ \cite{manore2014comparing}.

\section{Incidence Rates and Parameter Estimation}
Although the traditional approach when modeling disease dynamics is to assume the total population within a country is susceptible, we deviate from this convention. Specifically, we assume the susceptible population depends on the biology and binomics of the {\it{Aedes}} species as well as the country's geography and uniquely calculate the at-risk population (described in the following section). We then use the at-risk population as the size of the initial susceptible population within simulations.   While the deterministic model (\ref{ode}) captures the dynamics of the epidemic fairly well, we follow the opinion of the authors in \cite{elderd2006uncertainty}, namely that more attention should be paid to how uncertainty in parameter estimates might affect model predictions.  Thus, in Section \ref{bayes}, we focus on incorporating and quantifying the uncertainty of the disease process and the parameter values by means of embedding the deterministic model into a stochastic process.

\subsection{Calculated At-Risk Population}
The data utilized herein was reported by the Pan American Health Organization (PAHO) \cite{PAHOepicurve}, and is given by the number of cases, both confirmed and suspected, of Zika per week at the country level for Colombia, El Salvador and Suriname.  Simulations performed by using the entire country population as the number of initially susceptible humans mischaracterized the disease dynamics, leading to overestimates in the final size of an epidemic, see Appendix (Figure \ref{wayoff}).  Since dengue and Zika occur in the same areas, share a common vector and have similar asymptomatic rates, we calculate the at-risk population size per country for a Zika outbreak based on historical data for dengue from 1995 to 2015 within Colombia, El Salvador, and Suriname \cite{PAHO2}.  The year of highest incidence for dengue provides an approximation for the number of susceptible individuals in a fully na{\"i}ve at-risk population, which coincides with the dynamics of a newly emerging pathogen, such as Zika, that would spread rapidly in a completely susceptible population.  The World Health Organization (WHO) released a report on dengue \cite{world2007report}, stating ``...available results suggest that the actual number of cases of dengue may range from 3 to 27 times the reported [dengue] number."  Hence, we use a scalar multiple of the reported number of dengue cases during the highest incidence rate year to obtain a reasonable at-risk population count for the number of initial susceptible individuals with regards to a Zika epidemic.  Within each of the three countries, the at-risk population value is strictly less than the total country population.
%
%

Table \ref{denguedata} indicates the year with the highest incidence rate and the reported number of cases for that year in each country of interest.  See the Appendix (Figure \ref{incidencegraph}) for all historical data on dengue incidence rates for Colombia, El Salvador and Suriname.

\begin{table}[H]
\centering
\caption{Reported dengue values from WHO used to compute the at-risk population}
\label{denguedata}
\vspace{0.1in}
\begin{footnotesize}
\begin{tabular}{p{2cm}p{4cm}p{2cm}p{4cm}}
\hline
Country & Highest Incidence Rate & Year  & Reported Number of Cases\\
\hline
\multicolumn{4}{l}{{\color{white}Dependent variables / Populations}}\\
Colombia & $0.00342$ & $2010$ & \qquad$157,152$\\
El Salvador & $0.00875$ & $2014$ & \qquad$53,460$\\
Suriname & $0.00580$ & $2005$ & \qquad$2,853$\\
\\
\hline
\end{tabular}
\end{footnotesize}
\end{table}

We simulated Model (\ref{ode}) using the ode45 solver in MATLAB with chosen baseline parameter values (Table \ref{zikadata}) while assuming that a single infectious human exists at the start of the epidemic. The at-risk population size for Colombia is 2.75 times the number of reported dengue cases from 2010.  We estimated the multiplier for the at-risk population based on the best fit for the currently ongoing epidemic. The at-risk population size for El Salvador is 1.425 times the number of reported dengue cases from 2014, while the at-risk population size for Suriname is 7.75 times the number of reported dengue cases from 2005.  The initial size of the susceptible mosquito population is double the number of the at-risk population per country.   When starting our analysis with only one infected human, the simulated epidemic takes several weeks to ramp up to a detectable (reportable) level. If we begin simulations on the first day that a case is reported, the simulated peak occurs after the reported peak. Thus, we use shifted initial conditions (Table \ref{IC}) that correspond to the population state sizes obtained approximately on day 38, 118, and 50 from the original simulation using Model (\ref{ode}) of the epidemics in Colombia, El Salvador and Suriname, respectively.  We then repeat the process of simulating Model (\ref{ode}) using the shifted initial conditions to obtain solutions for the number of reported cases whose peak reporting weeks align more closely in time with that of the data. The precise values of the shifted initial conditions for each country, found in Table \ref{IC}, are of the form $[S_h,E_h,I_{r_h},I_h,R_h,S_v, E_v,I_v]$.

\begin{table}[t]
\centering
\caption{Computed At-Risk Population and Other Initial Conditions}
\label{IC}
\vspace{0.1in}
\begin{footnotesize}
\begin{tabular}{p{2cm}p{5cm}p{5cm}p{2cm}}
\hline
Country & Original Initial Conditions & Shifted Initial Conditions$^*$ & \qquad ~~$\psi$ \\
\hline
\multicolumn{4}{l}{{\color{white}Dependent variables / Populations}} \\
Colombia & $[432168,0,1,0,0,864336,0,0]$ & $[432163,1,0,1,4,864330,3,3]$&   \qquad 18\% \\
El Salvador & $[76181,0,1,0,0,152362,0,0]$ & $[75802,87,14,62,217,152082,145,135]$ &  \qquad18\%\\
Suriname & $[22111,0,1,0,0,44222,0,0]$ & $[22099,3,0,2,7,44212,5,5]$&  \qquad 22\%\\
\\
\hline
\end{tabular}

\hspace{-.15cm}$^*$Note: The shifted initial conditions were obtained approximately from Day 38, 118, and 50\\
\vspace{-.1cm}
\hspace{-.5cm}of the solution using the original initial conditions for Colombia, El Salvador and Suriname, respectively.
\end{footnotesize}
\end{table}

\begin{table}[H]
\centering
\caption{Ranges and Chosen Baseline Values for Model Parameters}
\label{zikadata}
\vspace{0.1in}
\begin{footnotesize}
\begin{tabular}{p{2cm}p{3cm}p{3cm}p{4cm}}
\hline
Parameter & Baseline & Range  & Reference\\
\hline
\multicolumn{4}{l}{{\color{white}Dependent variables / Populations}}\\
$1/\sigma_v$ & $3.85$ & $2 - 5.26$ & \cite{delatte2009influence,sivanathan2006ecology,manore2016defining}\\
$\beta_{hv} $& $0.54$ & $0.1 - 0.75$ & \cite{manore2014comparing,kucharski2016transmission,manore2016defining}\\
$1/\nu_h$ & $6$ & $3 - 12$ & \cite{campos2015zika,foy2011probable,fonseca2014first,kucharski2016transmission,lessler2016times,venturi2016autochthonous,WHOZika,manore2016defining}\\
$1/\gamma_h$ & $7$ & $3 - 14$ & \cite{campos2015zika,foy2011probable,fonseca2014first,kucharski2016transmission,lessler2016times,venturi2016autochthonous,WHOZika,manore2016defining}\\
$\psi$ & $0.18$ & $0.05 - 0.3$ & \cite{kucharski2016transmission,duffy2009zika,manore2016defining}\\
$\beta_{vh}$ & $0.75$ & $0.1 - 0.75$ & \cite{manore2014comparing,kucharski2016transmission,manore2016defining}\\
$1/\nu_v$ & $10.2$ & $4.5 - 17$ & \cite{kucharski2016transmission,wong2013aedes,andraud2012dynamic,boorman1956simple,manore2016defining}\\
$1/\mu_v$ & $18$ & $8 - 35$ & \cite{andraud2012dynamic,otero2006stochastic,fay1964biology,manore2016defining}\\
\\
\hline
\end{tabular}
\end{footnotesize}
\end{table}

\begin{figure}[H]
\centering
\subfigure[Colombia - The model initially fits the data well but was unable to capture the dynamics observed towards the end of the epidemic.]{\includegraphics[width=.45\textwidth]{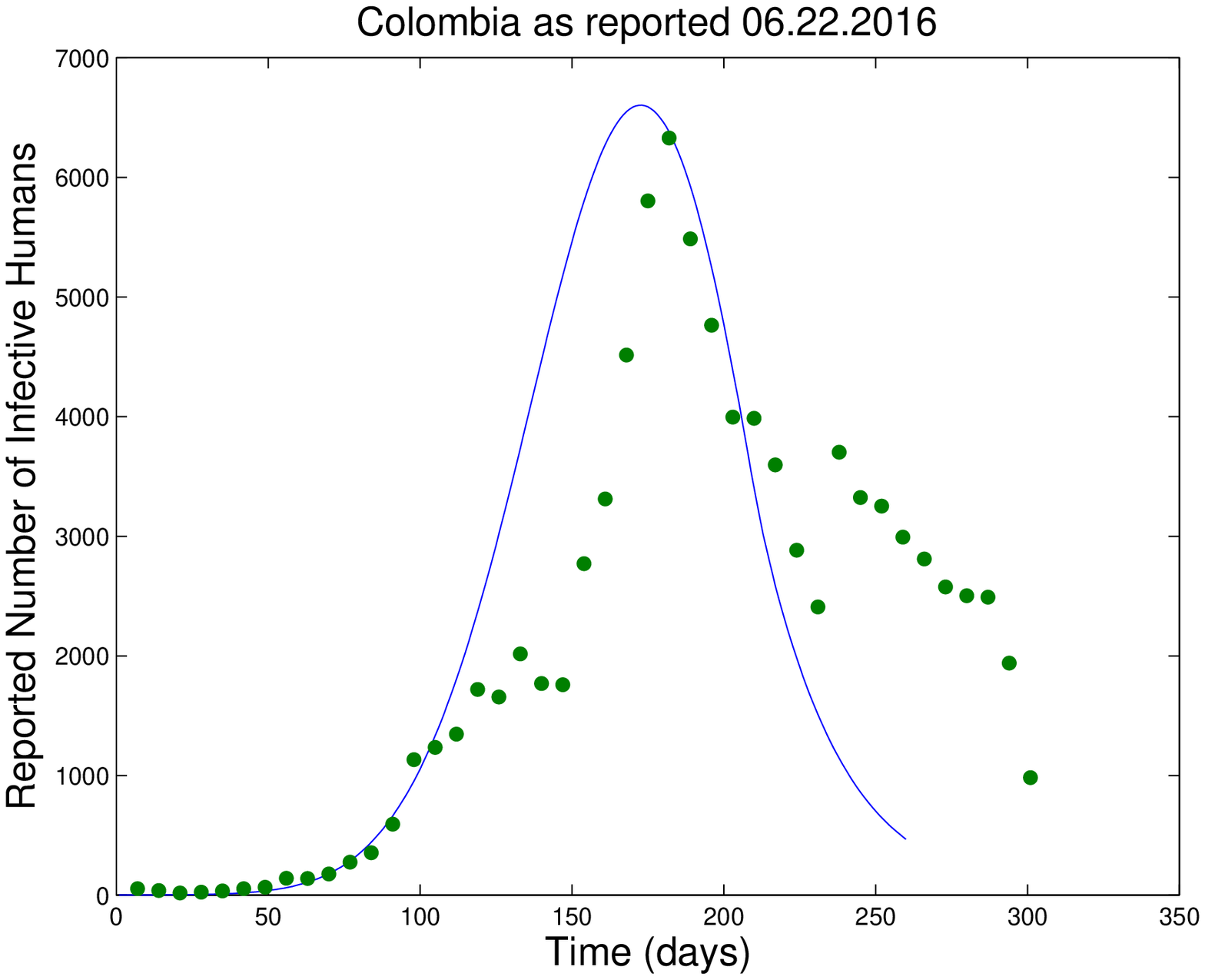}} \hspace{.2cm}
\subfigure[El Salvador - The model is able to capture the overall pattern of the epidemic but with a slightly later peak week.]{\includegraphics[width=.45\textwidth]{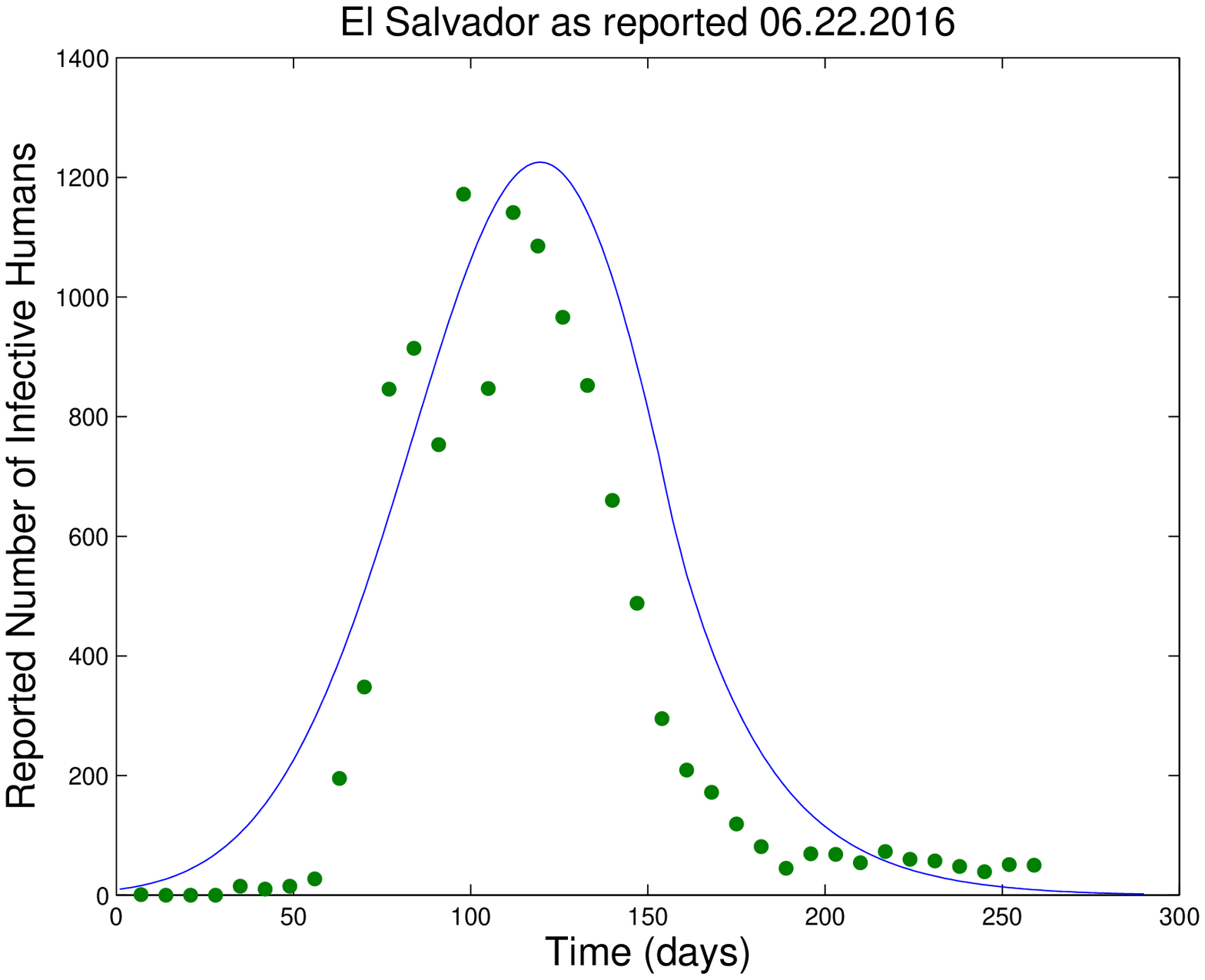}}
\subfigure[Suriname - The model is able to capture the overall pattern of the epidemic.]{\includegraphics[width=.45\textwidth]{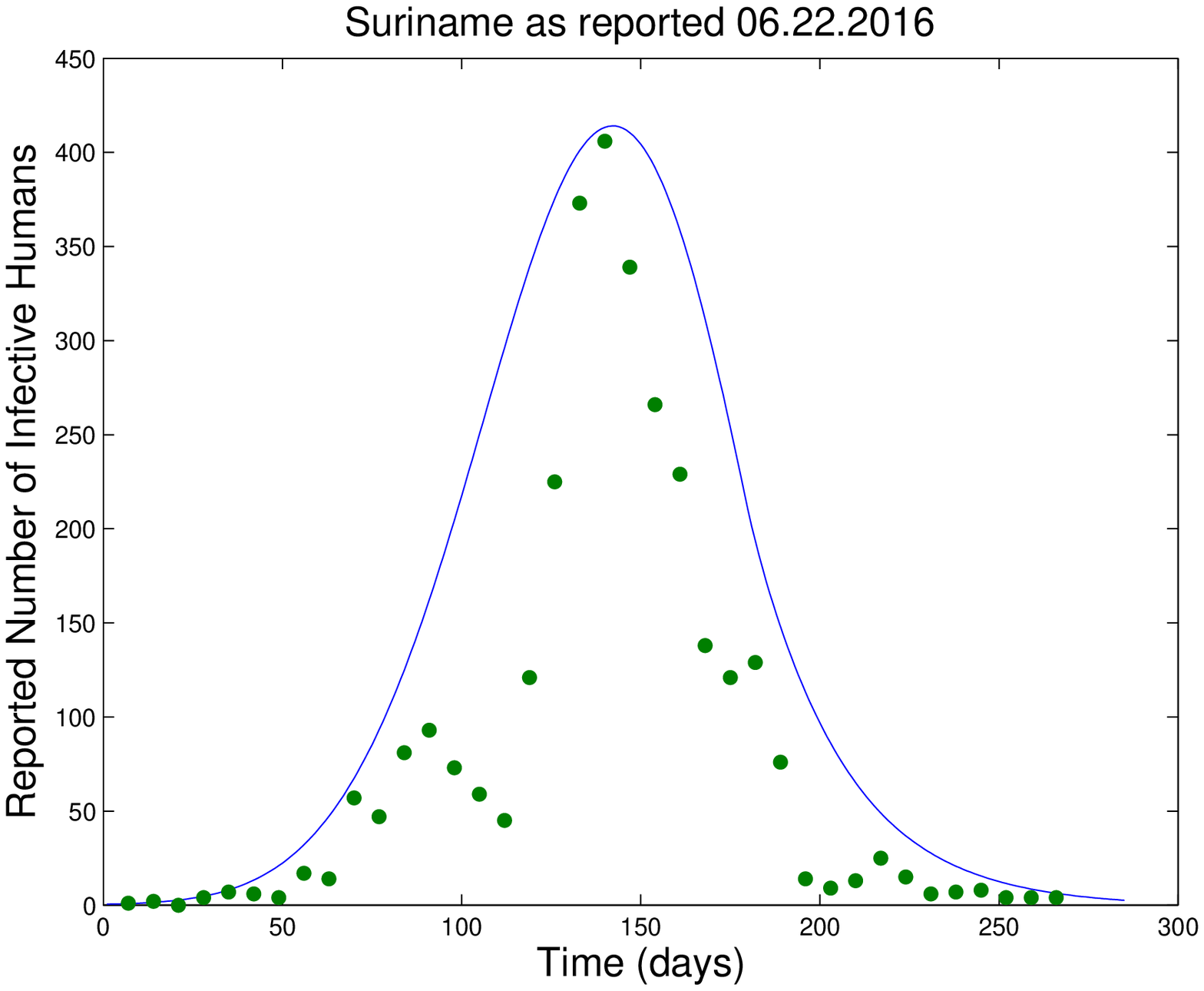}}
\caption{\footnotesize{Solutions of Model (\ref{ode}) are plotted in blue along with the weekly Zika data available from PAHO shown as green circles. These solutions used the shifted initial conditions (see Table \ref{IC}) and baseline parameters (see Table \ref{zikadata}). (a) Columbia: cases of Zika were reported first in Colombia on Epidemiological Week (EW) 32 of 2015.  (b) El Salvador: the first Zika cases were reported on EW 38 of 2015. (c) Suriname:  the reporting Zika cases began on EW 37 of 2015.}}
\label{eyeballmatch}
\end{figure}

The number of reported cases from our simulations, using the calculated at-risk population and shifted initial conditions, are compared to the PAHO Zika data in Figure \ref{eyeballmatch}.  The solutions, which are similar to the data in peak and general shape, identify reasonable initial conditions.  Although the parameter ranges considered are biologically reasonable, it is unknown which values within these ranges are the most accurate.  To obtain an expected value for a given parameter and describe the associated uncertainty within this quantity, we embed the deterministic Model (\ref{ode}) into a stochastic process enabling statistical inference.  This process is described in Section \ref{bayes}.

\subsection{An Embedded Stochastic Model}
\label{bayes}
Previous literature has provided biologically relevant parameter ranges (see Table \ref{zikadata}) for the model.  However, to obtain insight as to the distribution of these parameters across their ranges, we embed the deterministic system of ordinary differential equations (\ref{ode}) within a discrete-time stochastic process to obtain a corresponding stochastic model ({\ref{stoch}, \ref{stoch2}}) and perform an analysis in a Bayesian paradigm. This approach adds a probabilistic component to both the contact/infection process and the disease progression process by capturing uncertainty at each modeling stage, rather than just within the contact/infection process \cite{king2015avoidable}.  Therefore, embedding deterministic models into stochastic processes serves to (a) create a stochastic model informed by population-level dynamics which includes uncertainty in the entire disease process rather than just due to data collection, (b) more adequately capture uncertainty present in the modeling framework, which is particularly important in small to medium sized epidemics, and (c) provide a convenient framework for fast computation of model parameters.  The process by which we embed Model (\ref{ode}) into a stochastic process, \cite{mode2000stochastic}, is summarized in the following paragraphs.

Previously, we obtained specific rates for the transfer of both humans and mosquitos between their respective compartments, and these can be used to inform the stochastic analogues of these parameters.  Throughout, we impose that the new model be conservative, i.e. $S_h + E_h + I_{r_h} + I_h+R_h = N_h$ and $S_v + E_v +I_v = N_v$ where $N_h$ and $N_v$ are the total human and mosquito populations, respectively.  The assumption of conservation within the human population is plausible as the time span of the epidemic is much smaller than the average lifespan of a human.  We hold the mosquito population as constant for the convenience of calculations and analysis.
To approximate a continuous SEIR type model, we account for all events/transitions that may occur during time interval $(i, i+h_i]$ to be assigned to the related compartment on day $i$.  Thus, the rate of change for a given population size can be approximated by the difference between the previous and current time steps.  The model now takes the form

\begin{equation}
\label{stoch}
\left. \begin{aligned}
S_{{i+h_i}}&=S_{i}-E^*_{i+h_i}\\
E_{{i+h_i}}&=E_{i}+E_{i+h_i}^*-I_{i+h_i}^*\\
Ir_{i+h_i}&= Ir_{i} +\psi I_{i+h_i}^* - RIr_{i+h_i}^*  \\
I_{{i+h_i}}&=I_i+(1-\psi)I_{i+h_i}^*-RI_{i+h_i}^*\\
R_{i+h_i}&= R_{i} + RIr_{i+h_i}^* + RI_{i+h_i}^*\\
Sv_{{i+h_i}}&=Sv_i +dEv_{i+h_i}^*+ dIv_{{i+h_i}}^*- Ev_{i+h_i}^*\\
Ev_{{i+h_i}}&=Ev_{{i}} + Ev_{i+h_i}^* - Iv_{i+h_i}^* - dEv_{i+h_i}^*\\
Iv_{{i+h_i}}&=Iv_{i}+Iv_{{i+h_i}}^*-dIv_{{i+h_i}}^*.
\end{aligned}
\right \}
\end{equation}

%

\noindent We index time by $i$, and the temporal offset, $h_i$, which may not be constant but is known.   All quantities represent counts and quantities denoted by an asterisk represent transition counts. In regards to the terms $\psi I_{i+h_i}^*$ and $(1-\psi)I_{i+h_i}^*$, the parameter $\psi$ may yield values which are not integer counts.  To be more liberal with the reporting size we calculated the smallest integer not less than the corresponding value of $\psi I_{i+h_i}^*$. To be more conservative with under-reporting size we calculated the largest integer not greater than the corresponding value of $(1-\psi)I_{i+h_i}^*$.   When individuals can transition into a compartment via multiple routes, (e.g., recovered individuals can recover either with or without being reported as infected, $RIr_{i+h_i}^*$ and $RI_{i+h_i}^*$), two letters are used to denote the transition.  In these cases, the first letter denotes \textit{to which} compartment the individual transitions and the second denotes the compartment \textit{from which} the individual transitions.  The transition compartments which represent birth/death counts of the mosquito popultion on day $i+h_i$ are denoted, $dEv_{i+h_i}^*$ and $dIv_{{i+h_i}}^*$.  Note that these values are drawn based on the calculated population sizes at time $i+h_i$.  The compartments are labeled as follows: $S$ - susceptible humans, $E$ - latently infected humans, $Ir$ - infectious reported humans, $I$ - infectious unreported humans, $R$ - recovered humans, $Sv$ - susceptible mosquitoes, $Ev$ - latently infected mosquitoes, $Iv$ - infectious mosquitoes.
Finally, the stochastic components of the model are given by
\begin{equation}
\label{stoch2}
\begin{footnotesize}
\left. \begin{aligned}
E^*_{i+h_i}&\sim \hbox{Bin} \biggl (S_i,1-\exp(-\lambda_h h_i) \biggr), \quad
I_{i+h_i}^*\sim \hbox{Bin}\biggl (E_i,1-\exp(-\nu_h h_i) \biggr )\\
RIr_{i+h_i}^*&\sim  \hbox{Bin} \biggl (Ir_i, 1-\exp(-\gamma h_i) \biggr ), \quad
RI_{{i+h_i}}^*\sim \hbox{Bin} \biggl (I_{i},1-\exp(-\gamma h_i) \biggr )\\
Ev_{i+h_i}^*&\sim \hbox{Bin} \biggl (Ev_i,1-\exp(-\lambda_v h_i)\biggr), \quad
Iv_{i+h_i}^*\sim \hbox{Bin} \biggl (Iv_i,1-\exp(-\nu_v h_i)\biggr)\\
dEv_{i+h_i}^*&\sim \hbox{Bin} \biggl (Ev_{i+h_i},1-\exp(-\mu_h h_i)\biggr), \quad
dIv_{i+h_i}^*\sim \hbox{Bin} \biggl (Iv_{i+h_i},1-\exp(-\mu_h h_i)\biggr).
\end{aligned}
\right \}
\end{footnotesize}
\end{equation}\\


\section{ABC Algorithm and Computation}
\label{sim}
Previous investigations \cite{duffy2009zika,manore2014comparing,kucharski2016transmission,wong2013aedes} have established biologically accepted ranges for the parameters $\bf{\Theta} = [\sigma_v, \beta_{hv}, \nu_h, \psi, \beta_{vh}, \nu_v, \mu_v]$ used within \eqref{ode}, but the conversion of Model \eqref{ode} to Model (\ref{stoch}, \ref{stoch2}) will incorporate uncertainty into the disease process, and thus the values of these parameters.  The stochastic model (\ref{stoch}, \ref{stoch2}) also allows for Bayesian inference on parameter posteriors which have the form
$$f({\bf{\Theta}} | Y) \propto f(Y | {\bf{\Theta}})\pi({\bf{\Theta}})$$
\noindent where $f(Y | {\bf{\Theta}})$ is the stochastic data model, which for fixed values of ${\bf{\Theta}}$ can be used to generate the random epidemic process, and $\pi({\bf{\Theta}})$ is the prior distribution of the parameters.   Thus, given both the data model and the prior distribution, we are able to calculate the posterior distribution $f({\bf{\Theta}} | Y)$ up to a proportionality constant. This serves to update the distribution of the biologically accepted parameter ranges based on the actual epidemic data which we denote as $Y$.

The data model is constructed as a product of binomials with different sample sizes and probabilities across every time point. Determining a Maximum Likelihood Estimate (MLE) to provide a point estimate and standard deviation for parameters may provide estimates outside of the valid biological ranges.   In SEIR models, Markov Chain Monte Carlo (MCMC) methods produce parameter autocorrelations in chains which becomes problematic for tuning. We can avoid these obstacles by using Approximate Bayesian Computation (ABC).  This method was introduced by \cite{rubin1984bayesianly} to obtain an approximation of the true posterior distribution, $f({\bf{\Theta}} | Y)$.
ABC samples from the posterior by randomly selecting parameter values from the prior that could adequately generate the data.  In particular, random draws, ${\bf{\Theta^*}}$, from the prior distribution produce generated data sets, $X$, which are then compared to a given epidemic data set, $Y$, by means of a chosen distance metric, $\rho(X,Y)$.  Those values of ${\bf{\Theta^*}}$ that generate data sets which fit the given data will be accepted as valid draws from the posterior distribution, and this implicitly conditions the posterior on $Y$.  The algorithm of the computation is as follows, where $N$ is the total number of accepted generated data sets $X$:\\

For $j \leq N$
\begin{enumerate}
\item{Draw ${\bf{\Theta^*}} \sim {\text{Unif}}(a_k,b_k)$, where $(a_k,b_k)$ is the corresponding parameter range found in Table \ref{ABCranges} for all $k = 1,...,7$. }
\item{Generate $X$, time series data of the number of reported infectives, from Model (\ref{stoch}, \ref{stoch2})}
\item{Calculate fitness of data using $\rho(X,Y)$}
\item{Set ${\bf{\Theta}}_{[j]} \leftarrow \bf{\Theta^*}$ if $\rho(X,Y) \leq \epsilon$  and set $j \leftarrow j+ 1$ else return to Step $1$.  }
\end{enumerate}

\begin{table}[H]
\centering
\caption{Ranges for Uniform Distributions}
\label{ABCranges}
\vspace{0.1in}
\begin{footnotesize}
\begin{tabular}{p{4cm}p{2cm}p{.01cm}p{.01cm}}
\hline
Parameter & Range & \\
\hline
\multicolumn{4}{l}{{\color{white}Dependent variables / Populations}} \\
$1/\sigma_v$& $0.5 - 10$ & & \\
$\beta_{hv}$& $0.01 - 1$ & & \\
$1/\nu_h$& $2 - 20$ & & \\
$1/\gamma_h$& $7$ & & \\
$\psi$& $0.05 - 0.35$ & & \\
$\beta_{vh}$& $0.01 - 1$ & & \\
$1/\nu_v$& $4 - 20$ & & \\
$1/\mu_v$& $7 - 50$ & & \\
\\
\hline
\end{tabular}\\
Note: We expand the ranges given in Table \ref{zikadata} as these are educated approximations and\\
\vspace{-.1cm}
\hspace{-1.1cm}there is uncertainty with regards to the true range of most of these parameters.
\end{footnotesize}
\end{table}

\begin{figure}[htp]
\centering
\subfigure[Colombia]{\includegraphics[height=.35\textwidth]{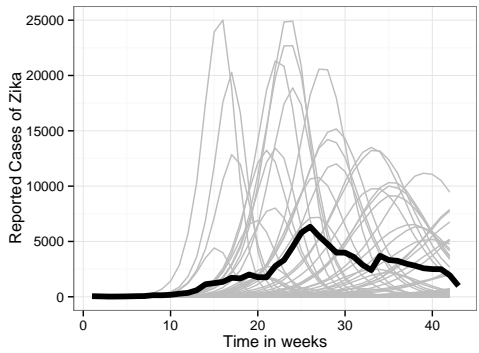}}
\hspace{0.2in}
\subfigure[El Salvador]{\includegraphics[height=.35\textwidth]{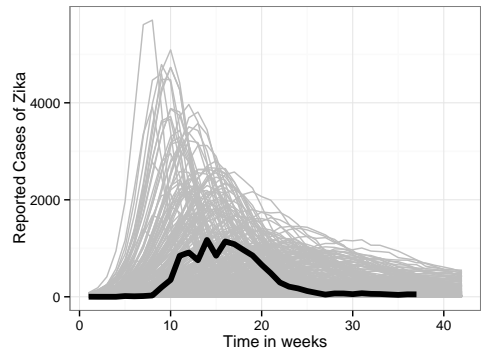}}
\subfigure[Suriname]{\includegraphics[height=.35\textwidth]{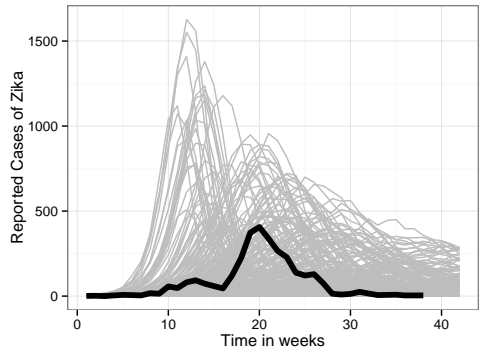}}
\caption{\footnotesize Each subfigure contains the results of using randomly drawn parameter vectors in Model (\ref{stoch}, \ref{stoch2}) to generate outcomes of the epidemic in which a metric is not applied. For all three countries the biologically accepted parameter values generate epidemics that vary greatly from the characteristics of the observed epidemics, particularly in the peak and duration.}
\label{random}
\end{figure}

\begin{figure}[H]
\centering

\subfigure[Colombia]{\includegraphics[height=.35\textwidth]{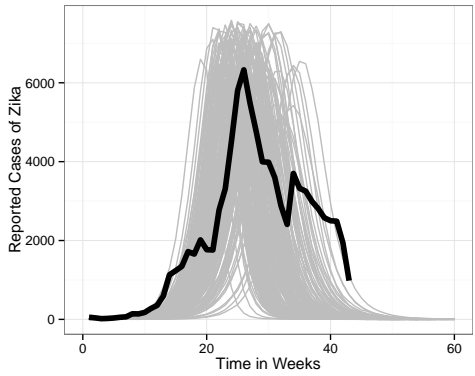}}
\subfigure[El Salvador]{\includegraphics[height=.35\textwidth]{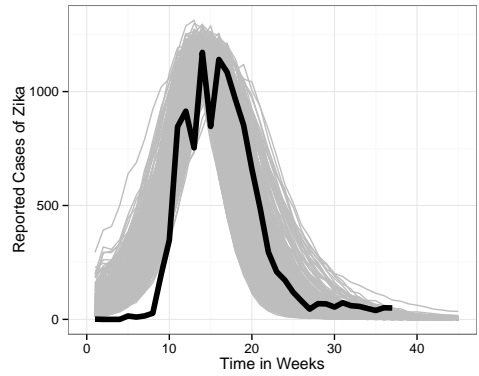}}
\subfigure[Suriname]{\includegraphics[height=.35\textwidth]{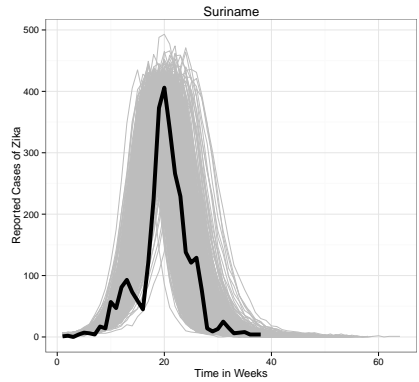}}
\label{accepted}
\caption{\footnotesize Each subfigure compares the stochastic model epidemic outcomes from 500 of the 10,000 accepted randomly generated parameter vector draws (gray) and the observed epidemic in each country (black). The envelope size is $(1/20,20)$ for Colombia (a) with the occurrence of peak week within $\pm4$ weeks of the known peak week and the peak number of new infections less than or equal to a 20\% deviation from the true value. The envelope size for El Salvador (b) is $(2/3,2)$ with the occurrence of peak week within $\pm2$ weeks of the known peak week and the peak number of new infections less than or equal to a 5\% deviation from the true value.  The data from Suriname (c) was compared to generated data with an envelope size of $(1/4,4)$ and the occurrence of peak week within $\pm2$ weeks of the known peak week and the peak number of new infections less than equal to a 5\% deviation from the true value.}
\end{figure}

The metric, $\rho(X,Y)$, is considered a distance between the generated data set, $X$, and the observed data set, $Y$, typically based on sufficient statistics of the parameter space ${\bf{\Theta}}$.  If $\rho(X,Y)$ is small enough, i.e., $\rho(X,Y) \leq \epsilon$ for some fixed small value $\epsilon > 0$, then
$$f({\bf{\Theta}} |\rho(X,Y) \leq \epsilon ) \approx f({\bf{\Theta}} | Y).$$
One often uses sufficient statistics in defining $\rho(X,Y)$. These are statistics for which the data distribution conditioned on the sufficient statistic is free of ${\bf{\Theta}}$ (e.g., the sufficient statistic contains all information about ${\bf{\Theta}}$ that the full data set contains).  By definition, the full data set is a sufficient statistic for ${\bf{\Theta}}$.  Since a sufficient statistic of lower dimension than the full data set cannot be readily computed for our model, we consider a pointwise envelope metric comparing point by point the $L_1$-norm at every time step.  Thus, if every data point of $X$ and $Y$ is close, then it directly follows that any statistics computed from $X$ and $Y$, including sufficient statistics, will also be close.  For the ABC method, the exact posterior distribution of  ${\bf{\Theta}}$ can be found by accepting the simulated data set in which $X = Y$.  This is computationally infeasible, however, so we instead consider $X$, such that  $\frac{X_t}{Y_t} \in (\epsilon_1,\epsilon_2)$ for all time steps $t$ with a corresponding peak week and epidemic duration to $Y$, as an acceptable epidemic.  We call $(\epsilon_1,\epsilon_2)$ the \textit{envelope of tolerance} around the observed data set, $Y$, which is a common method for assessing a stochastic SEIR model fit \cite{lekone2006statistical,porter2013path, porter2016spatial}. In Figure \ref{random}, the observed data set is compared to randomly generated data sets without calculating a metric for validation of the drawn parameter values.  Figure \ref{random}  shows that many randomly drawn and biologically accepted values of ${\bf{\Theta}}$ generate epidemic outcomes which have distinctly different characteristics than the observed epidemic.  Conversely, Figure \ref{accepted} demonstrates that epidemic outcomes of the accepted ${\bf{\Theta}^*}$ values generate epidemics with similar characteristics to that of the observed epidemic in regards to the total number of new infections each day, peak week occurrence, and peak value.  Figure \ref{accepted}(b) and \ref{accepted}(c) show generated epidemics similar to the data in both peak and duration.  Note that this does not occur in Figure \ref{accepted}(a).  The accepted parameter values obtained using the ABC method yield outbreaks which vary greatly in peak occurrence, leading us to conclude that the dynamics of the country-level data are different than those in the deterministic model. Therefore the ABC method does not generate infectious curves similar to the Colombia data. Figure \ref{accepted}(a) was generated from acceptances using an envelope of $(1/20,20)$.  A tighter envelope of $(1/10,10)$ was computed; however, the method ran for approximately 35 days to obtain the same number of acceptances as the $(1/20,20)$ envelope with no visible changes in the same plot as Figure \ref{accepted}(a).  Because acceptances for tighter envelopes are not possible, we can conclude the model is improperly specified to capture the dynamics of Colombia at the country level. Reasons for this outcome are discussed in Section \ref{conclusions}.

\begin{figure}[H]
\centering
\subfigure[El Salvador - $\psi$]{\includegraphics[height=.35\textwidth]{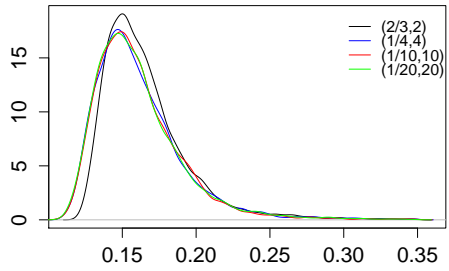}}
\subfigure[El Salvador - $\mathscr{R}_0$ ]{\includegraphics[height=.35\textwidth]{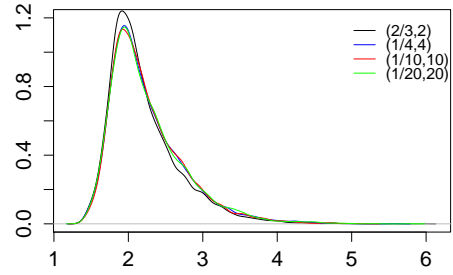}}
\subfigure[Suriname - $\psi$]{\includegraphics[height=.35\textwidth]{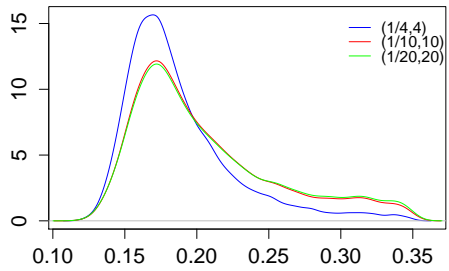}}
\subfigure[Suriname - $\mathscr{R}_0$]{\includegraphics[height=.35\textwidth]{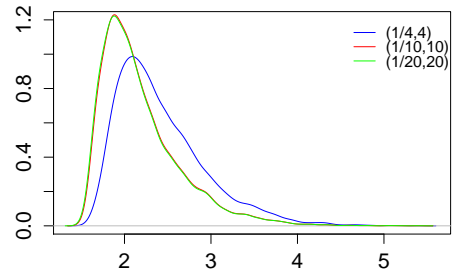}}
\caption{\footnotesize Subfigures (a) and (c) display the kernel density plots of the proportion of reported infectious, $\psi$, with various envelope sizes for El Salvador and Suriname, respectively, while subfigures (b) and (d) show the kernel density plots of the reproductive number, $\mathscr{R}_0$ for El Salvador and Suriname, respectively.  As the legends indicate, an envelope size of $(2/3,2)$ is displayed by the black curve, an envelope size of $(1/4, 4)$ is displayed in blue, the red curve represents an envelope size of $(1/10,10)$, and similarly the green curves indicate an envelope size of $(1/20,20)$ using green. Note that envelopes of all four sizes were able to generate acceptances from the El Salvador data while the data from Suriname only allowed for acceptance to be generated from envelope sizes of $(1/4,4)$, $(1/10,10)$, and $(1/20,20)$.}
\label{kernels}
\end{figure}

Figure \ref{kernels} depicts the effect of generating accepted epidemics with smaller and smaller envelopes of tolerance on the posterior distributions.  We plot the kernel densities of both the reporting rate, $\psi$, and reproductive number, $\mathscr{R}_0$, distributions for El Salvador and Suriname.  While only two parameters are shown here, the same observations are found in the kernel density plots of the other parameters of interest.  We see in Figure \ref{kernels} that envelopes of various size produce similar but different kernel densities, with the tighter envelopes producing a more peaked density distribution.  A slight shift in the peak of the kernel densities is observed in Figures \ref{kernels}(a) and \ref{kernels}(d).  This would indicate the need to shrink the envelope further to obtain an estimation that is closer to the true posterior distribution.  However, attempts to generate data sets satisfying tighter envelopes could not produce acceptances, indicating that these envelope sizes are the best possible fits of the stochastic $SEI_rIRS_vE_vI_v$ model to the observed country-level data.  Further analysis would require refining the dynamics of Model (1) and Model (2,3) or using data at a more refined spatial scale.


\section{Results}
\label{results}

\begin{figure}[H]
\centering
\subfigure[El Salvador - $\psi$]{\includegraphics[width=.3\textwidth]{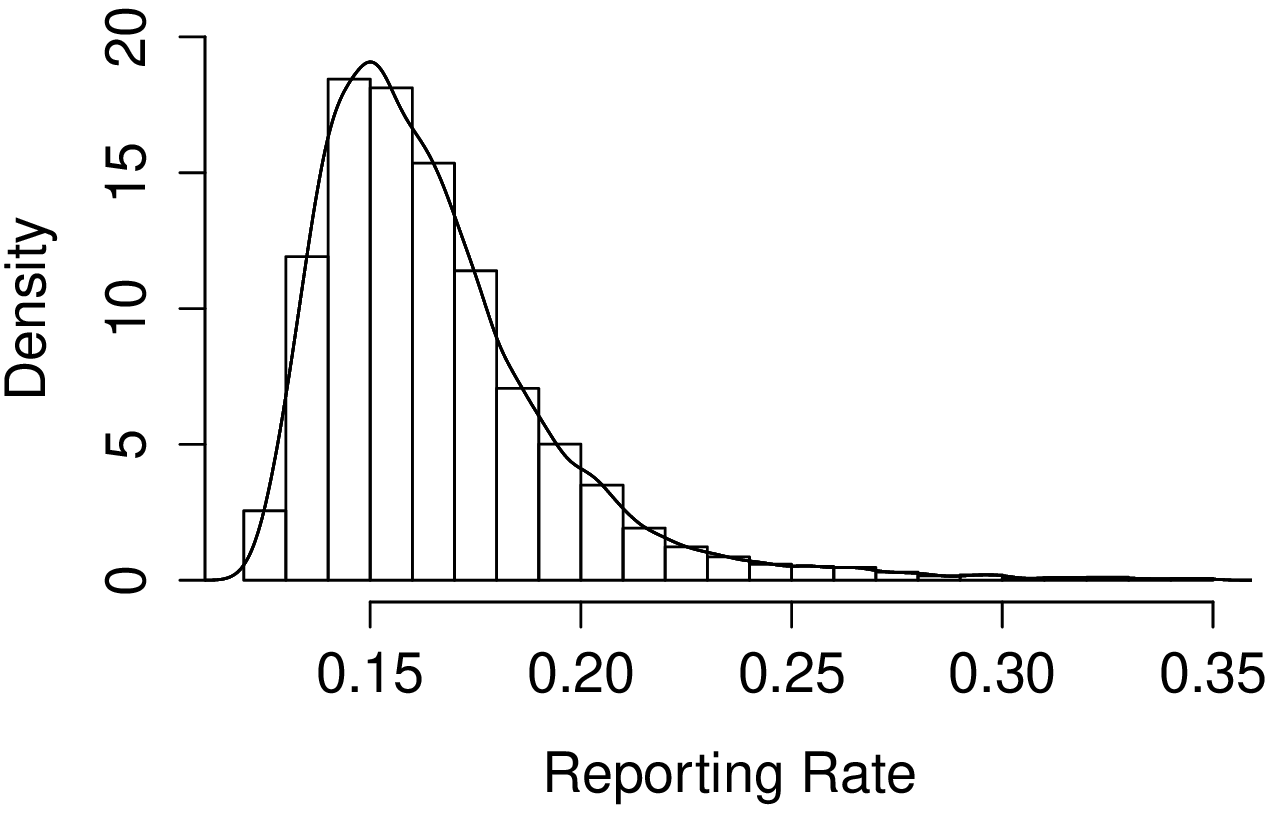}}
\subfigure[El Salvador - $\sigma_v\beta_{hv}$]{\includegraphics[width=.3\textwidth]{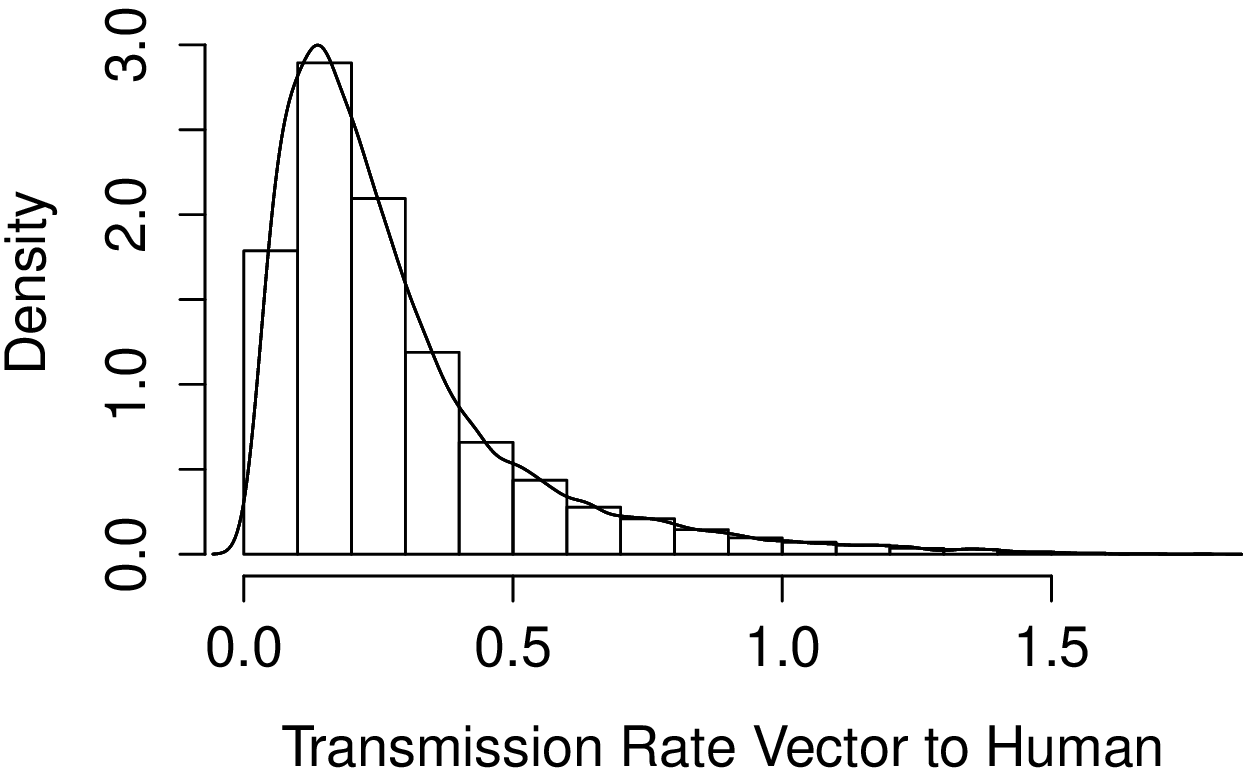}}
\subfigure[El Salvador - $\sigma_v\beta_{vh}$]{\includegraphics[width=.3\textwidth]{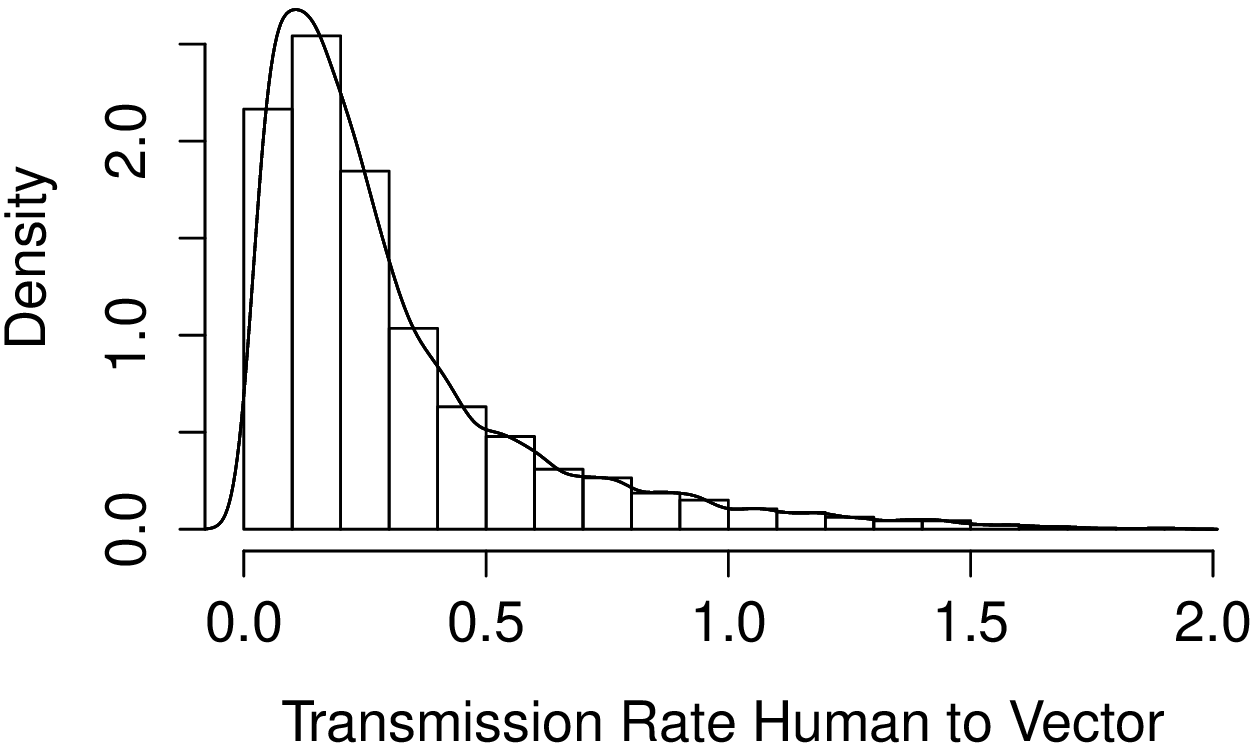}}
\caption{ABC generated histograms and kernel density plots for the reporting rate, $\psi$, transmission rate for the human population, $\sigma_v\beta_{hv}$, and mosquito population, $\sigma_v\beta_{vh}$ in El Salvador.}
\label{ABChistElSal}
\end{figure}

\begin{figure}[H]
\centering
\subfigure[Suriname - $\psi$]{\includegraphics[width=.3\textwidth]{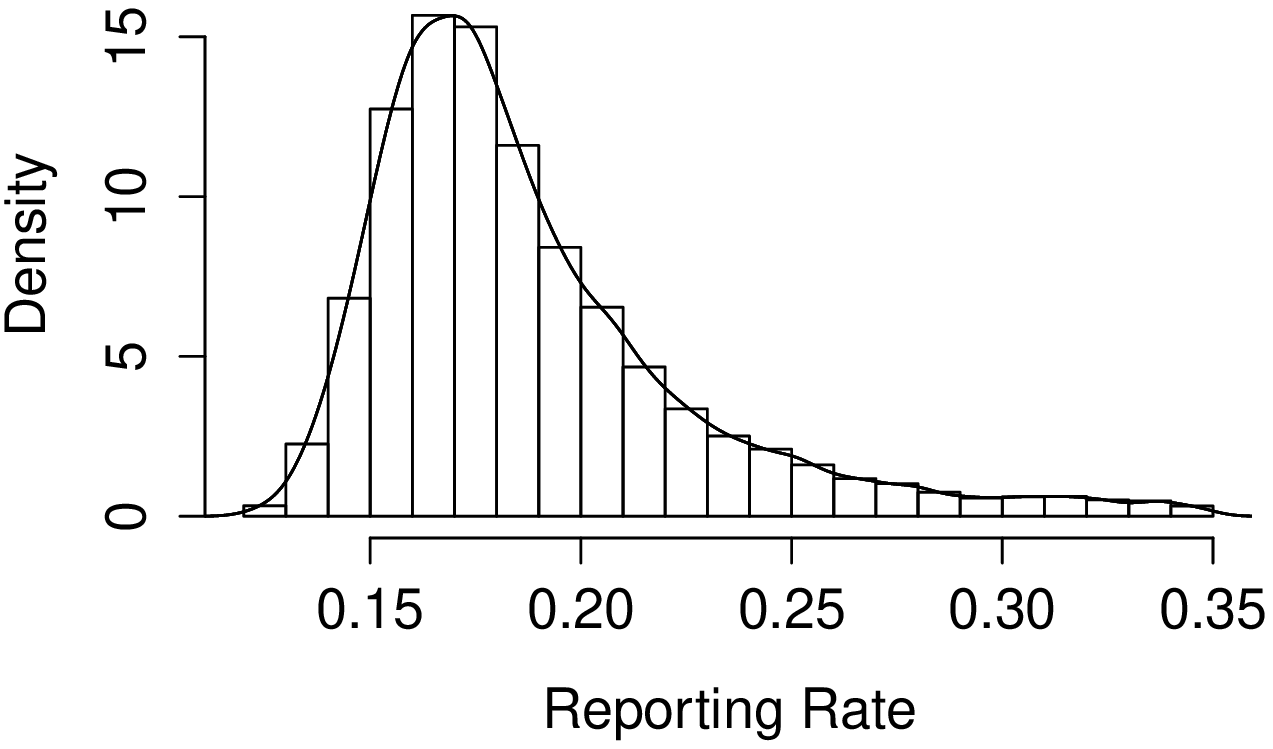}}
\subfigure[Suriname - $\sigma_v\beta_{hv}$]{\includegraphics[width=.3\textwidth]{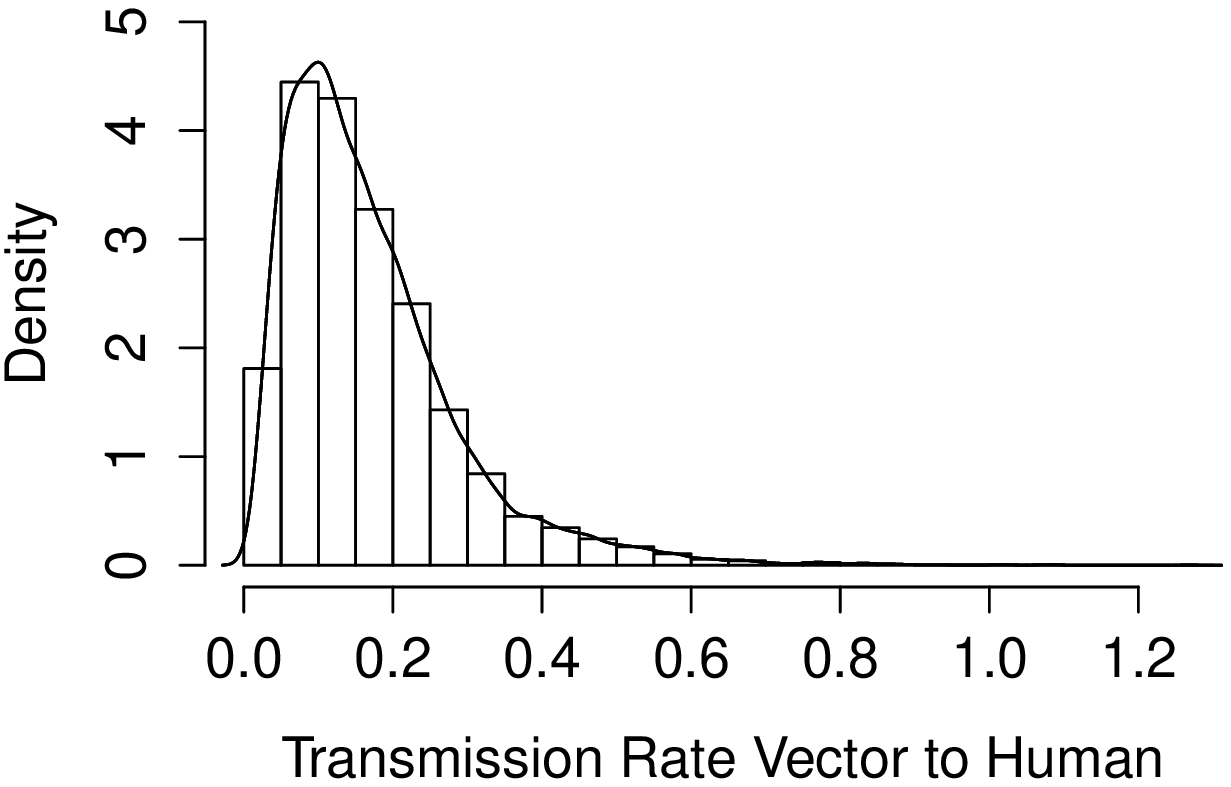}}
\subfigure[Suriname - $\sigma_v\beta_{vh}$]{\includegraphics[width=.3\textwidth]{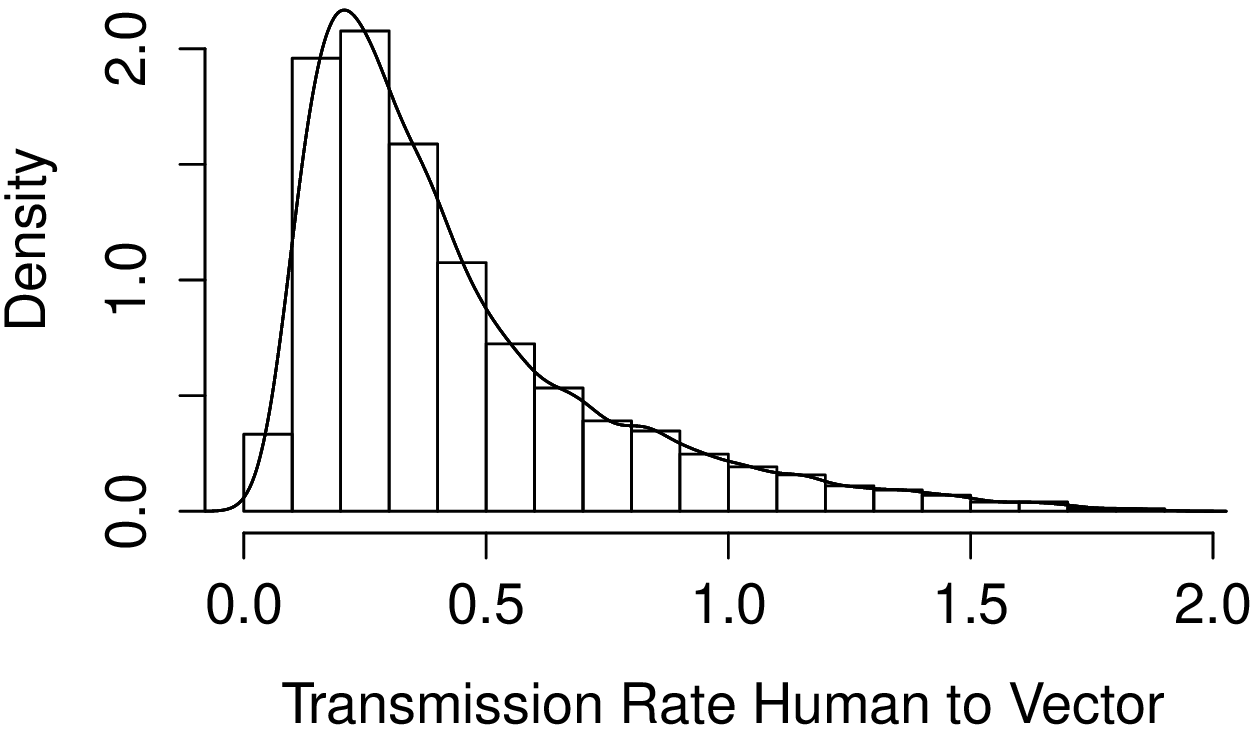}}
\caption{ABC generated histograms and kernel density plots for the reporting rate, $\psi$, the force of infection for the human population, $\sigma_v\beta_{hv}$, and mosquito population, $\sigma_v\beta_{vh}$ in Suriname.}
\label{SuriABChist}
\end{figure}

\begin{figure}[H]
\centering
\subfigure[El Salvador - $\mathscr{R}_0$]{\includegraphics[width=.45\textwidth]{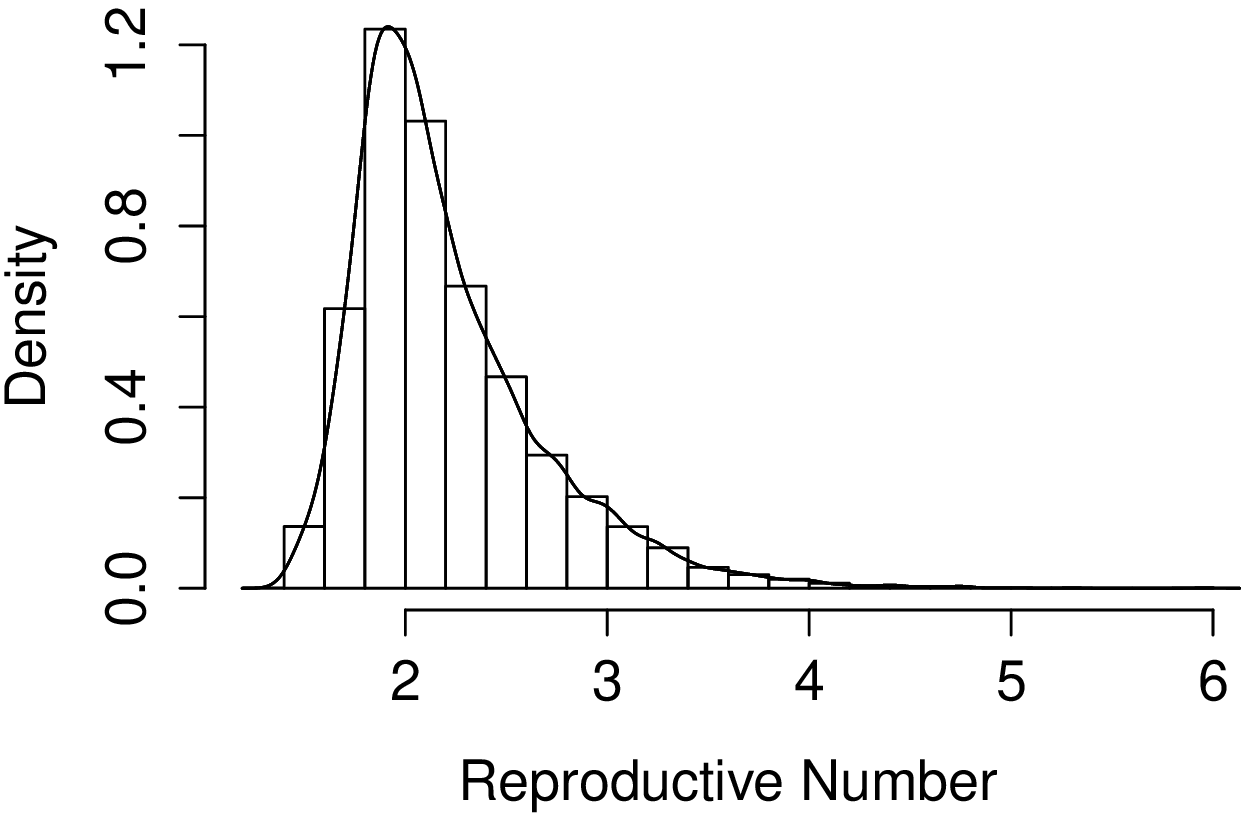}}
\subfigure[Suriname - $\mathscr{R}_0$]{\includegraphics[width=.45\textwidth]{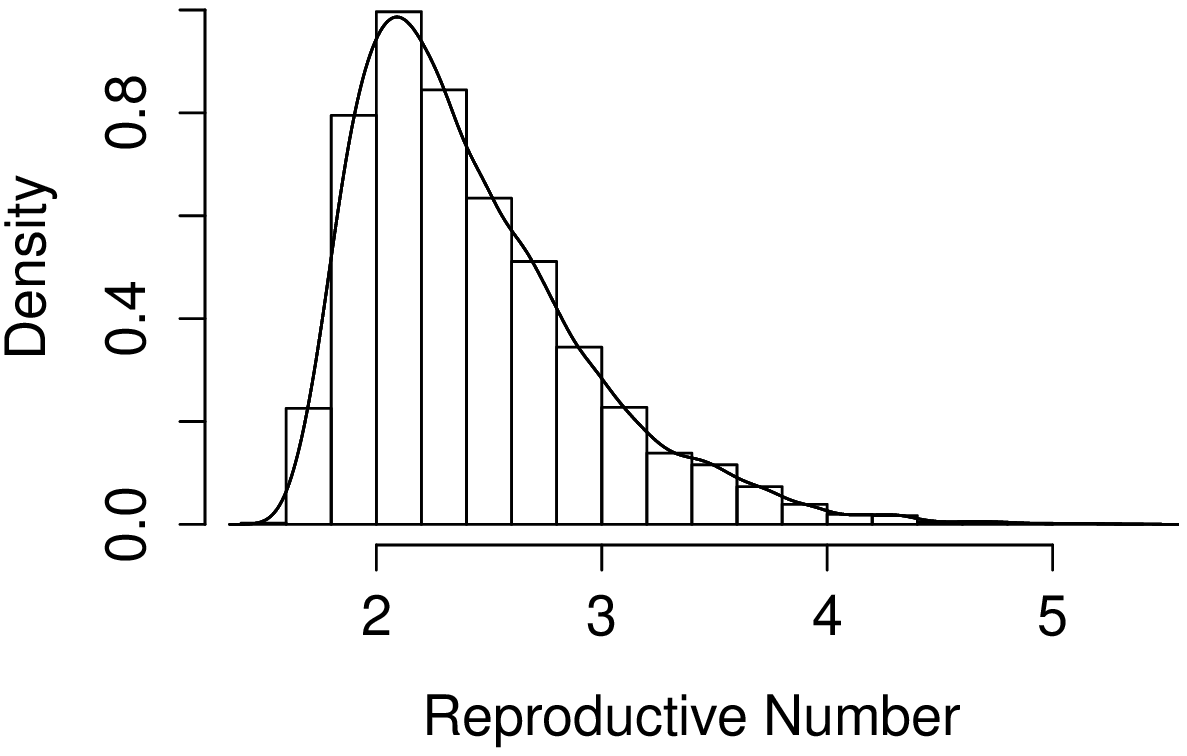}}
\caption{ABC generated histograms and kernel density plots of the Reproductive Number for El Salvador (a) and Suriname (b).}
\label{histrepro}
\end{figure}

\begin{figure}[H]
\centering
\subfigure[El Salvador]{\includegraphics[width=.45\textwidth]{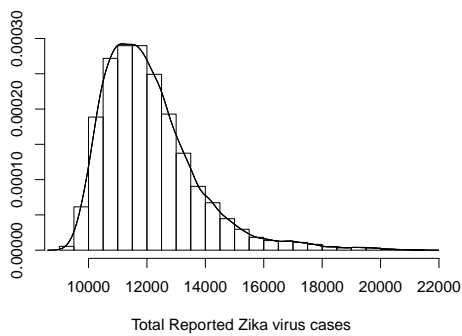}}
\subfigure[Suriname ]{\includegraphics[width=.45\textwidth]{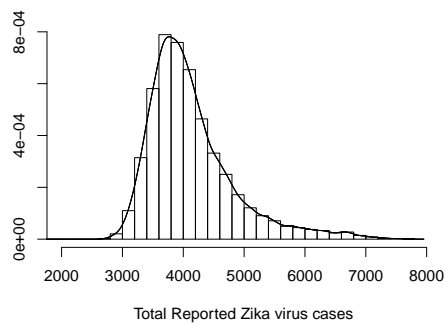}}
\caption{Histogram with kernel density plot of the total number of reported Zika virus cases generated from the ABC accepted parameter values for El Salvador (a) and Suriname (b).}
\label{reported}
\label{hist_infectives}
\end{figure}

\begin{figure}[H]
\centering
\subfigure[El Salvador]{\includegraphics[width=.45\textwidth]{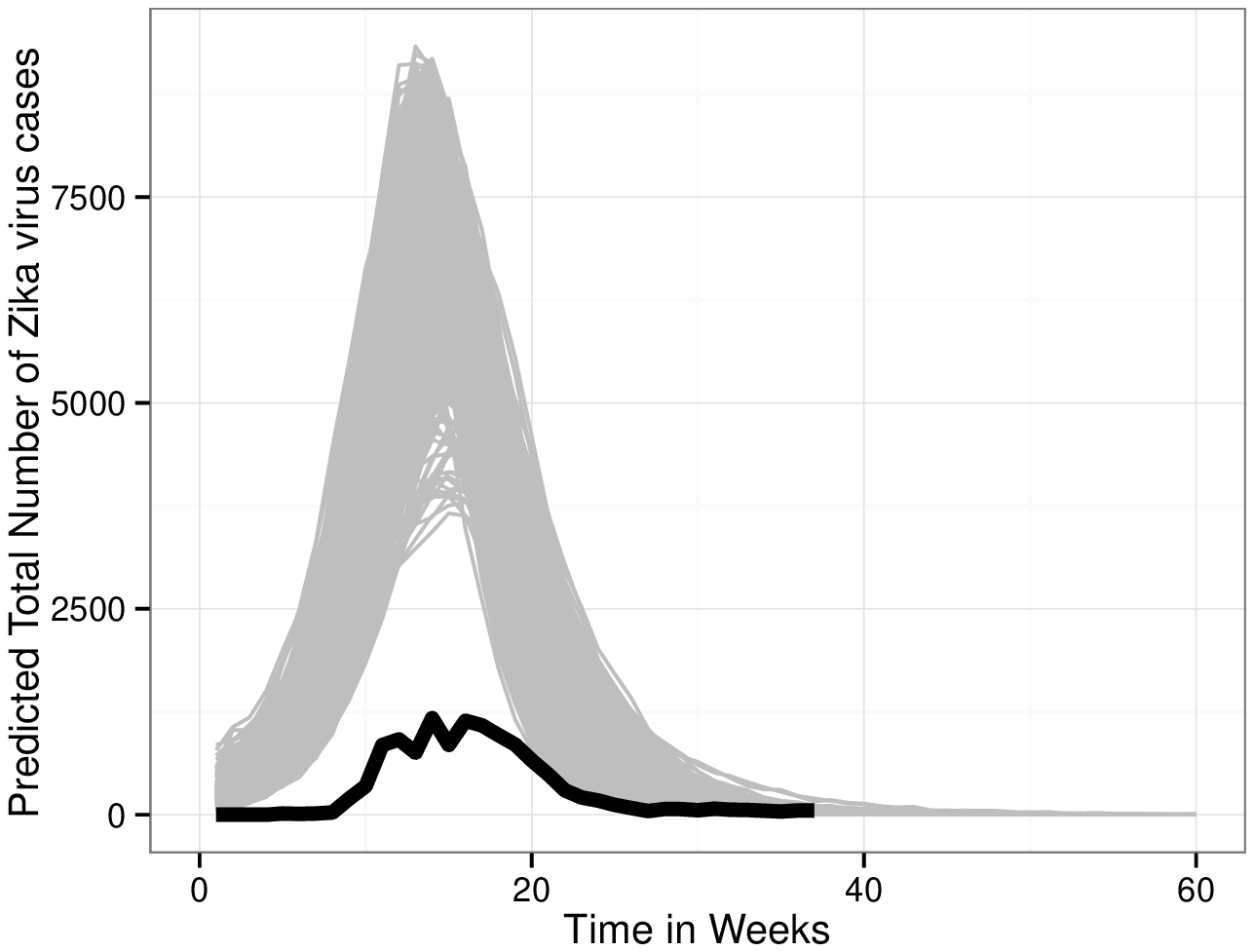}}
\subfigure[Suriname ]{\includegraphics[width=.45\textwidth]{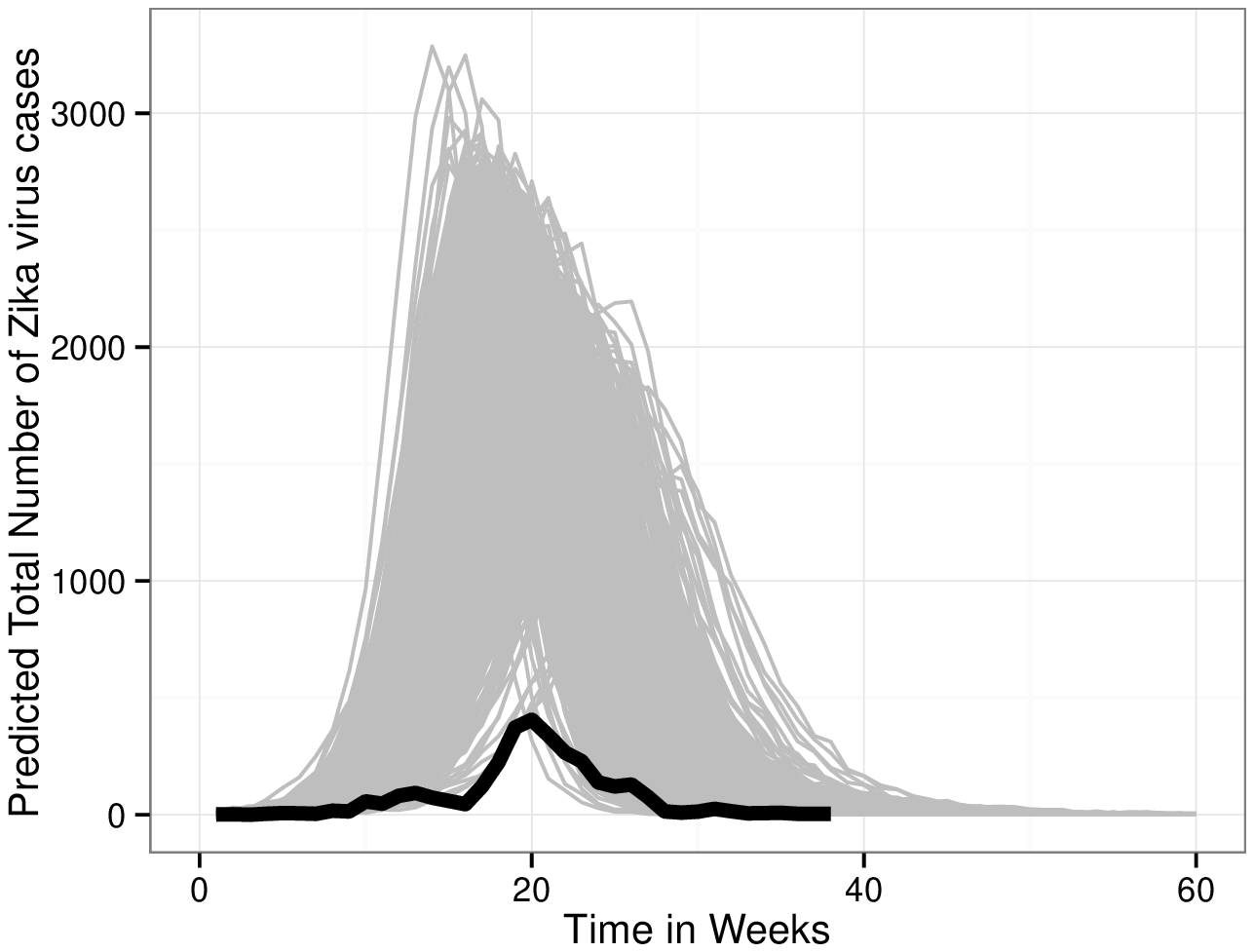}}
\caption{Generated time series data (grey) for the total number of Zika cases (reported + unreported) from the ABC-accepted parameter values as compared to the observed data (black) reported by PAHO for El Salvador (a) and Suriname (b).}
\label{timeseriesinfectives}
\end{figure}

\begin{figure}[H]
\centering
\subfigure[El Salvador]{\includegraphics[width=.45\textwidth]{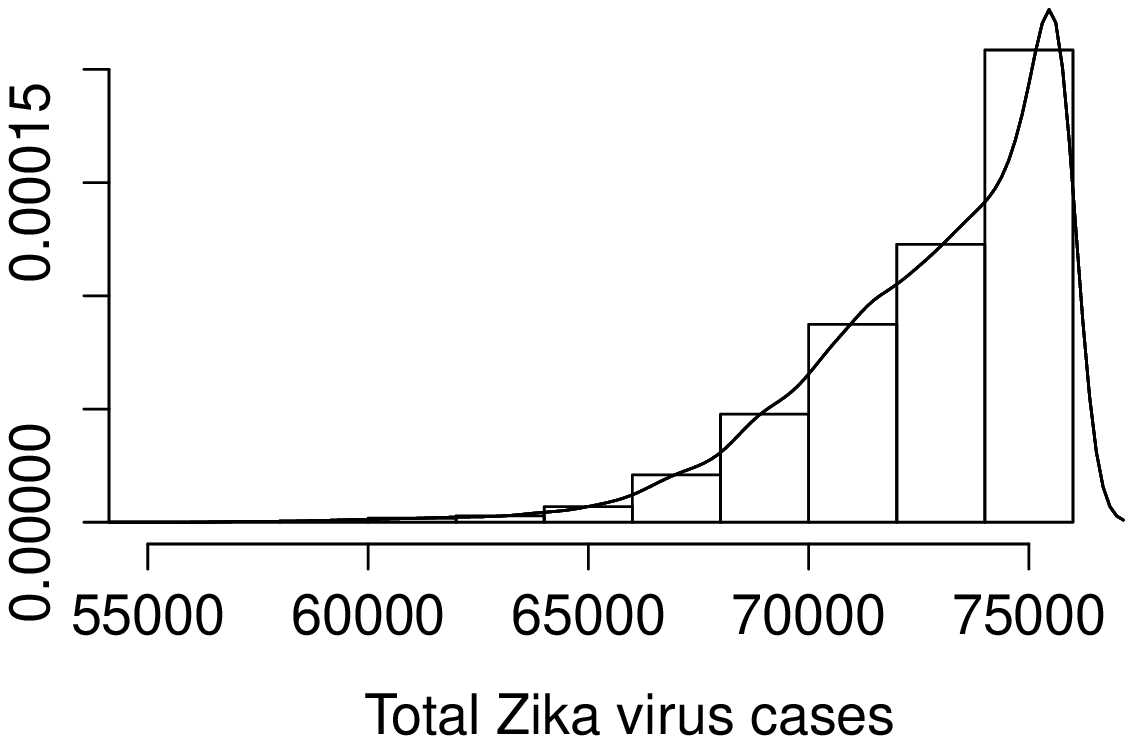}}
\subfigure[Suriname ]{\includegraphics[width=.45\textwidth]{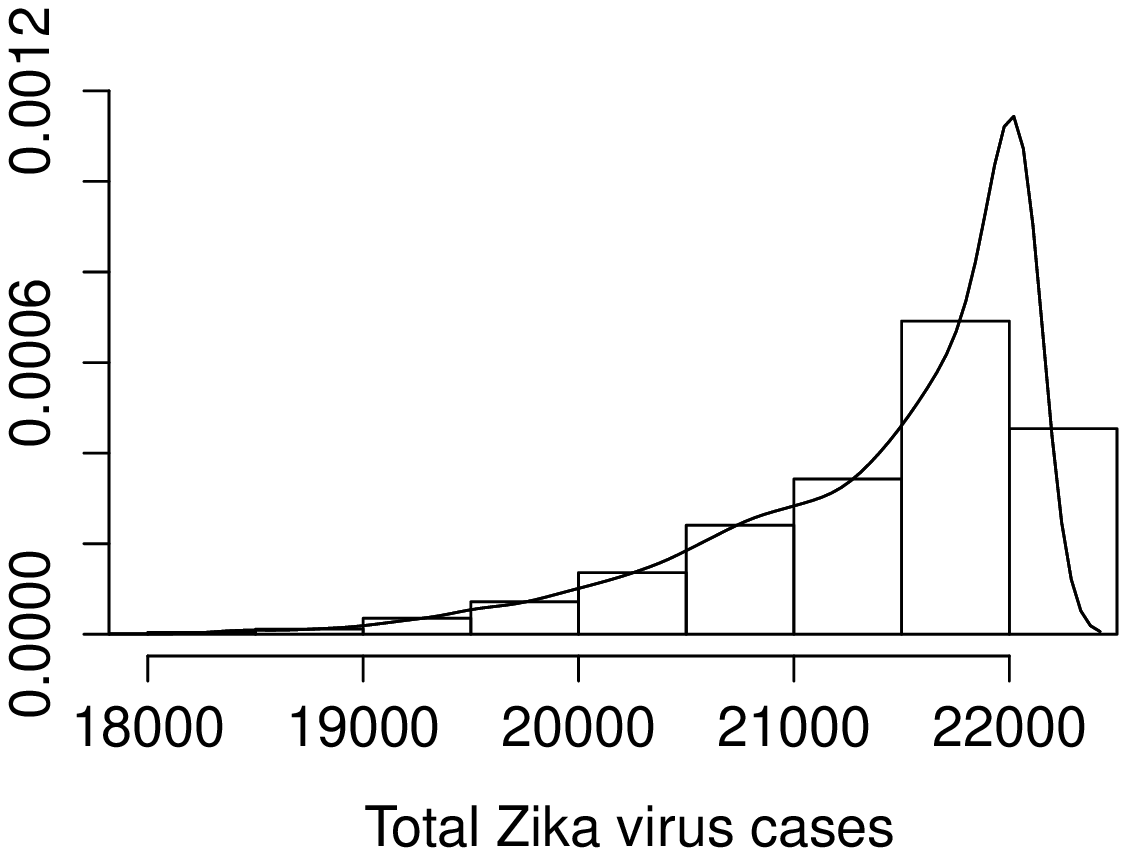}}
\caption{Histogram and kernel density plot of generated total number of Zika virus cases, both reported and unreported, from the ABC accepted parameter values for El Salvador (a) and Suriname (b).}
\label{gen_total_infectives}
\end{figure}

We assigned uniform distributions to the accepted biological ranges found in Table \ref{ABCranges}. After 10,000 acceptances from the ABC algorithm, the histograms and kernel densities for selected parameters from El Salvador are given in Figures \ref{ABChistElSal} and \ref{histrepro}.  Selected parameters from Suriname are given in Figures \ref{SuriABChist} and \ref{histrepro}.  Figure \ref{reported} displays the number of reported cases from the generated epidemics based on 10,000 accepted parameter values. The time series of the number of total cases during the generated epidemics of both reported and unreported infectives is found in Figure \ref{timeseriesinfectives}.  The histogram and kernel density plots of the total number of Zika virus cases, reported and unreported are shown in Figure \ref{gen_total_infectives}.



\begin{table}[H]
\centering
\caption{Selected Parameter Statistics for El Salvador}
\label{stats1}
\vspace{0.1in}
\begin{footnotesize}
\begin{tabular}{p{2cm}p{3cm}p{3cm}p{3cm}}
\hline
& ~~$\psi$ &~~ $\sigma_v\beta_{hv}$ & ~~$\sigma_v\beta_{vh}$  \\
\hline
\multicolumn{4}{l}{{\color{white}Dependent variables / Populations}} \\
mean & $0.1654$ & $0.2808$&  $0.3053$ \\
median & $0.1593$ & $0.2124$ &  $0.2130$\\
mode &$0.1501$ &$0.1372$ &$0.1091$\\
95\% C.I. &$[0.1248, 0.2205]$&$[0.0191, 0.7782]$&$[0.0119, 0.9244]$\\
\\
\hline
\end{tabular}
\end{footnotesize}
\end{table}

\begin{table}[H]
\centering
\caption{Selected Parameter Statistics for Suriname}
\label{stats2}
\vspace{0.1in}
\begin{footnotesize}
\begin{tabular}{p{2cm}p{3cm}p{3cm}p{3cm}}
\hline
& ~~$\psi$ &~~ $\sigma_v\beta_{hv}$ & ~~$\sigma_v\beta_{vh}$  \\
\hline
\multicolumn{4}{l}{{\color{white}Dependent variables / Populations}} \\
mean & $0.1877$ & $0.1705$&  $0.4352$ \\
median & $0.1777$ & $0.1429$ &  $0.3365$\\
mode &$0.1693$ &$0.0998$ &$0.2069$\\
95\% C.I. & $[0.1304,0.2717]$ & $[0.0165,0.4054]$&  $[0.04287,1.1241]$\\
\\
\hline
\end{tabular}
\end{footnotesize}
\end{table}

\begin{table}[H]
\centering
\caption{Reproductive Number Statistics per Country}
\label{R0}
\vspace{0.1in}
\begin{footnotesize}
\begin{tabular}{p{4cm}p{3cm}p{3cm}}
\hline
\quad~~$\mathscr{R}_0$  & \quad El Salvador & \quad Suriname \\
\hline
\multicolumn{3}{l}{{\color{white}Dependent variables / Populations}} \\
\quad~~mean & \quad $2.2055$ & \quad $2.4225$ \\
\quad~~median & \quad $2.0854$ & \quad $2.3062$ \\
\quad~~median$^2$& \quad $4.3487$ & \quad $5.3186$\\
\quad~~mode &\quad $1.9174$ & \quad $2.0901$\\\
\quad~$95\%$ C.I. & [1.4664, 3.1658] & [1.6526, 3.4704]\\
\\
\hline
\end{tabular}\\
\hspace{-.15cm}Note: The median$^2$ is an approximation of the type of reproduction number (i.e. human-to-human)\\
\vspace{-.1cm}
\hspace{-.5cm}as opposed to $\mathscr{R}_0$, which is the single generation reproduction number.
\end{footnotesize}
\end{table}
%

\begin{table}[H]
\centering
\caption{Statistics for the Generated Total Cases (Reported and Unreported) per Country}
\label{R0}
\vspace{0.1in}
\begin{footnotesize}
\begin{tabular}{p{4cm}p{3cm}p{3cm}}
\hline
El Salvador  &  Reported Cases & Total Cases \\
\hline
\multicolumn{3}{l}{{\color{white}Dependent variables / Populations}} \\
\quad~~mean & \quad $12107$ & \quad $72721$ \\
\quad~~median & \quad $11810$ & \quad $73395$ \\
\quad~~mode &\quad $11158$ & \quad $75457$\\\
\quad~$95\%$ C.I. & [9606,15321] & [67317, 75889] \\
\quad~~PAHO data & \quad11825\\
\\
\hline
Suriname & Reported Cases & Total Cases \\
\hline
\multicolumn{3}{l}{{\color{white}Dependent variables / Populations}} \\
\quad~~mean & \quad $4132$ & \quad $21390$ \\
\quad~~median & \quad $3975$ & \quad $21647.5$ \\
\quad~~mode &\quad $3758$ & \quad $22018$\\\
\quad~$95\%$ C.I. & [3013, 5697] & [19853,221027]\\
\quad~~PAHO data & \quad3042\\
\\
\hline
\end{tabular}
\end{footnotesize}
\end{table}

\section{Conclusions and Discussion}
\label{conclusions}
Within the present study, a new stochastic model was formulated to describe the spread of the Zika virus within Colombia, Suriname, and El Salvador. The variability in per capita susceptibility within each country was introduced by uniquely calculating the at-risk population based on historical data for dengue virus, whose epidemic characteristics are similar to that of Zika.  The initial population state sizes were fit to this data in order to create similar values of weekly reported cases of Zika, both suspected and confirmed, to those reported by the PAHO using baseline values of parameters to evaluate Model (\refeq{ode}).  Once the at-risk population was estimated, the initial population state sizes for these nations were fixed in order to estimate parameter values and obtain more informative distributions than the accepted biological parameter ranges.  The deterministic system (\refeq{ode}) was then embedded into a stochastic process to obtain a more general stochastic model (\ref{stoch}, \ref{stoch2}).  For each of the three nations, an ABC algorithm was implemented using (\ref{stoch}, \ref{stoch2}) to compute approximate posterior distributions of the parameters conditioned on the data.  By obtaining these posterior distributions, the uncertainty in parameter values for each country can be quantified by way of informative statistics, such as the mean, mode, median and variance, to more accurately describe rates within the system.

Properties of the disease within El Salavador and Suriname were accurately described by the model.  El Salvador is estimated to have a mean reporting rate of $16.5\%$ which is near the upper bound of the previous estimate of \cite{kucharski2016transmission} with a credible interval of $[12.5\%, 22\%]$.  The mean values of the forcing terms, $\lambda_h$ and $\lambda_v$, are approximately $0.28$ and $0.31$ with credible intervals of $[0.0191, 0.7782]$ and $[0.0119, 0.9244]$, respectively.  Suriname has a mean reporting rate greater than the predicted interval of \cite{kucharski2016transmission} at $18.8\%$ with a credible interval of $[13\%, 27\%]$. The forcing terms of Suriname, $\lambda_h$ and $\lambda_v$, had mean values of $0.17$ and $0.43$ with credible intervals of $[0.0165,0.4054]$ and $[0.0429,1.1241]$, respectively.
The basic reproductive number, $\mathscr{R}_0$, was defined such that ${\mathscr{R}_0}^2$ yields the number of secondary human infections within a fully susceptible population arising from a single new human infection.  In Suriname, the mean value of ${\mathscr{R}_0}^2$ was $5.31$, while the mean reproductive number was $4.35$ in El Salvador.  These quantities are similar to other predictions for the mean of ${\mathscr{R}_0}^2$ found in \cite{nishiura2016transmission,towersestimation}.

Though this analysis has provided additional insight into the spread of the disease, our methods were unable to accurately estimate the aforementioned statistics for Colombia (see Section \ref{sim}) as the ABC method revealed a poor fit for the data obtained from this nation.  One possible reason for this poor fit could be the appearance of a second peak in the epidemic, see Figure \ref{eyeballmatch}(a).  While there is a distinct and large peak during EW 26, a smaller second peak occurs during EW 34.  This second peak may result from reporting error as this data was in no way cleaned once obtained from the PAHO website.  Another possible explanation is a second outbreak in a disjoint location from the first outbreak.  In particular, the extreme topographical variations within Colombia could lead to a delay in the spread of the disease across the entire country.  Thus, we conclude that the epidemiological characteristics of Zika in Colombia must be studied at a more granular spatial level, differentiating amongst regions or even counties and cities.

We also estimated the total number of infected people in El Salvador to be 72,721 (with an estimated 12,107 reported cases) and the total number of people infected in Suriname to be 21,390 (with an estimated 4,132 reported cases). So, about 95\% of the at-risk (high-risk) populations were infected by the end of the outbreak. The predicted \textit{reported} country-level incidence for El Salvador is 0.0019 and for Suriname is 0.0073. However, the \textit{true} country-level incidence for El Salvador is predicted to be 0.0119 while for Suriname it is 0.0396 assuming total populations of 6,117,145 and 540,612, respectively. From a public health perspective, our model indicates that about 6 times as many people were infected than were reported in El Salvador and 5 times as many were infected than were reported in Suriname. Depending on the percent of the at-risk population that was pregnant during the outbreak, our model suggests that a larger number of birth defects than indicated by the reported number of cases could be expected in these countries. Interestingly, although the values for the mosquito extrinsic incubation period and mosquito lifespan had relatively wide ranges even among mean, median, and mode values, the probability of a mosquito surviving the incubation period (Tables 12 and 13) was quite similar across statistics and countries at about 60\%. This corresponds to the extrinsic incubation period lasting around 2/3 of the average mosquito lifespan. A major difference between predicted parameter values for El Salvador and Suriname occurs in the transmission probabilities. While the values for $\beta_{hv}$ and $\beta_{vh}$ for El Salvador were similar (both estimated to be about 0.42), in Suriname, the mosquito-to-human probability of transmission, $\beta_{hv}$ was consistently less than half that of the human-to-mosquito transmission, $\beta_{vh}$, with a median of 0.21 and 0.56, respectively. Since these terms capture many intrinsic uncertainties in the transmission process, it is hard to interpret the meaning of this difference. It could be an artifact of the model or could indicate a reduced efficiency of the mosquitoes in Suriname in passing on the virus.

In conclusion, our research provides important parameter estimates for the spread of Zika in El Salvador and Suriname, along with uncertainty quantification and credible intervals for those parameters. We estimated the basic and type reproduction numbers and the total number of people infected - quantities needed to inform assessments of economic cost and risk, among other factors. We found that the type reproduction number is higher in Suriname than in El Salvador, indicating a higher risk in Suriname. This could be explained by differences in climate between the two countries or in other socio-economic or geographic factors affecting mosquito-borne disease transmission. Additionally, our methods could be applied to other countries or regions experiencing outbreaks to estimate region-specific parameters and provide decision makers with important information about surveillance and control both at present and in the future. Another advantage of this method is that it can indicate what scale is appropriate for these calculations. For example, we found that a country-level analysis of the Colombia data was not appropriate. In the future, it would be interesting to apply the model to a regional Colombia data set.

In future studies, the initial population state sizes obtained in the calculations of the respective at-risk populations could be considered parameters themselves. Hence, one could perform a statistical analysis on acceptable ranges for such initial values to quantify the uncertainty of these populations. For instance, this could be done by generating a posterior distribution for the initial susceptible population for each country, as well as the parameters in the system.  Additionally, the mosquito population for each country was assumed to be constant for convenience.  However, utilizing a more realistic model of the total mosquito population which changes in time (similar to the methods in \cite{manore2014comparing}) may yield different results, and this would also be a suitable direction for future research. Finally, the recovery period, $1/\gamma_h$, was held constant in the current study due to the assumption that its mean value had been medically established.  Still, another investigation using the methods developed herein, but considering a uniformly-distributed prior for the recovery period, may provide better insight into the distribution and expected value of this period. We conclude that additional studies are needed to fully understand Zika virus transmission dynamics. However, our research suggests that the reporting rates in El Salvador and Suriname are quite low and thus, additional surveillance systems may be needed to measure the true burden of Zika in these countries.

%
%
%

\section{Acknowledgments}
This work was supported by NSF SEES grant CHE - 1314029, NSF RAPID (DEB 1641130), and NSF EDT grant DMS-1551229. SP was partially supported by NSF under grant DMS-1614586.  SD was partially supported by NIH/NIGMS/MIDAS under grant U01-GM097658-01. LANL is operated by Los Alamos National Security, LLC for the Department of Energy under contract DE-AC52-06NA25396. Approved for public release: LA-UR-17-20963. The funders had no role in study design, data collection and analysis, decision to publish, or preparation of the manuscript.

\section{Appendix}
\subsection{Results of Model {\ref{ode}} using total country populations as the initial size susceptible populations}
\begin{figure}[H]
\centering
\subfigure[Colombia]{\includegraphics[width=.35\textwidth]{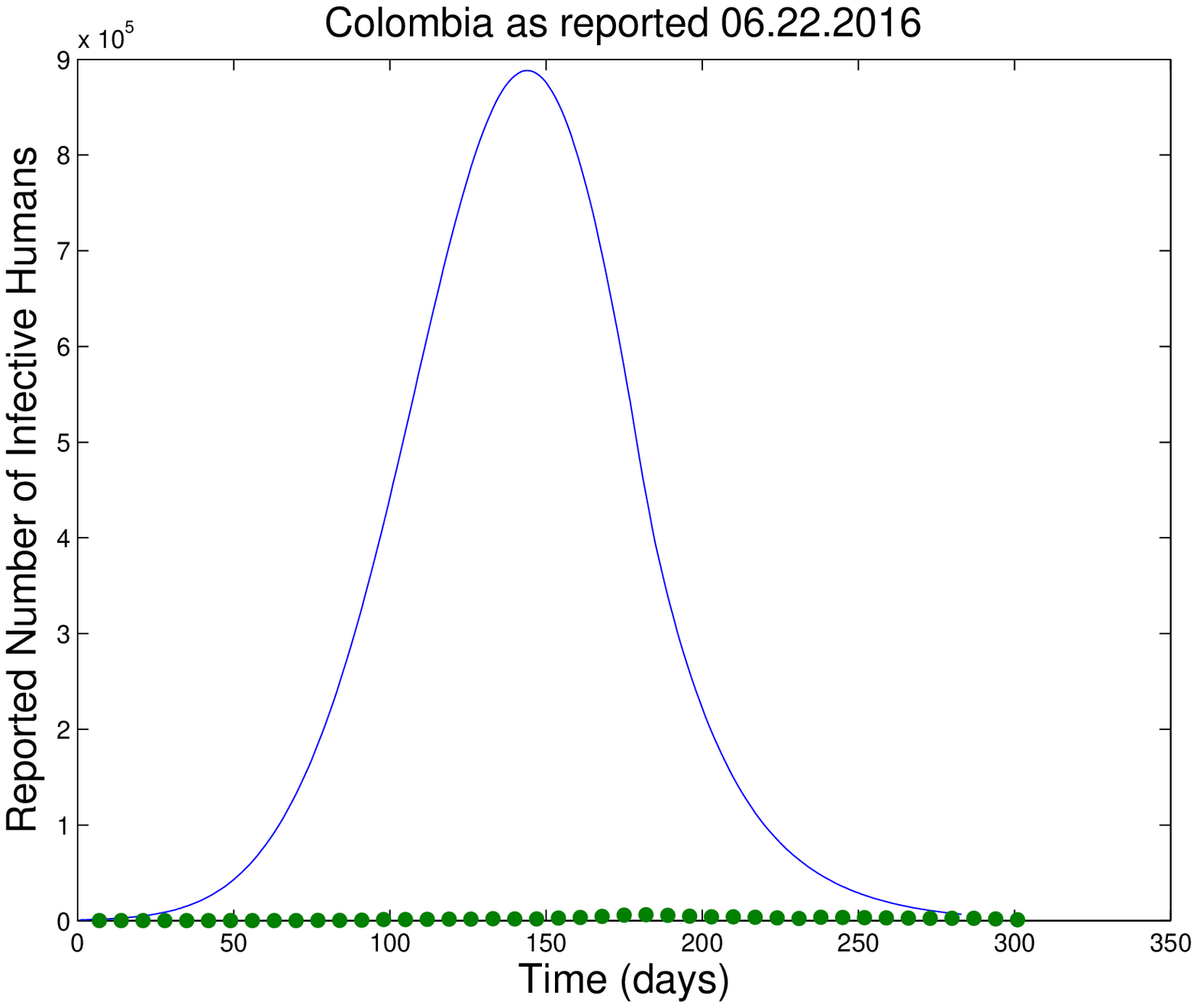}}
\subfigure[El Salvador]{\includegraphics[width=.35\textwidth]{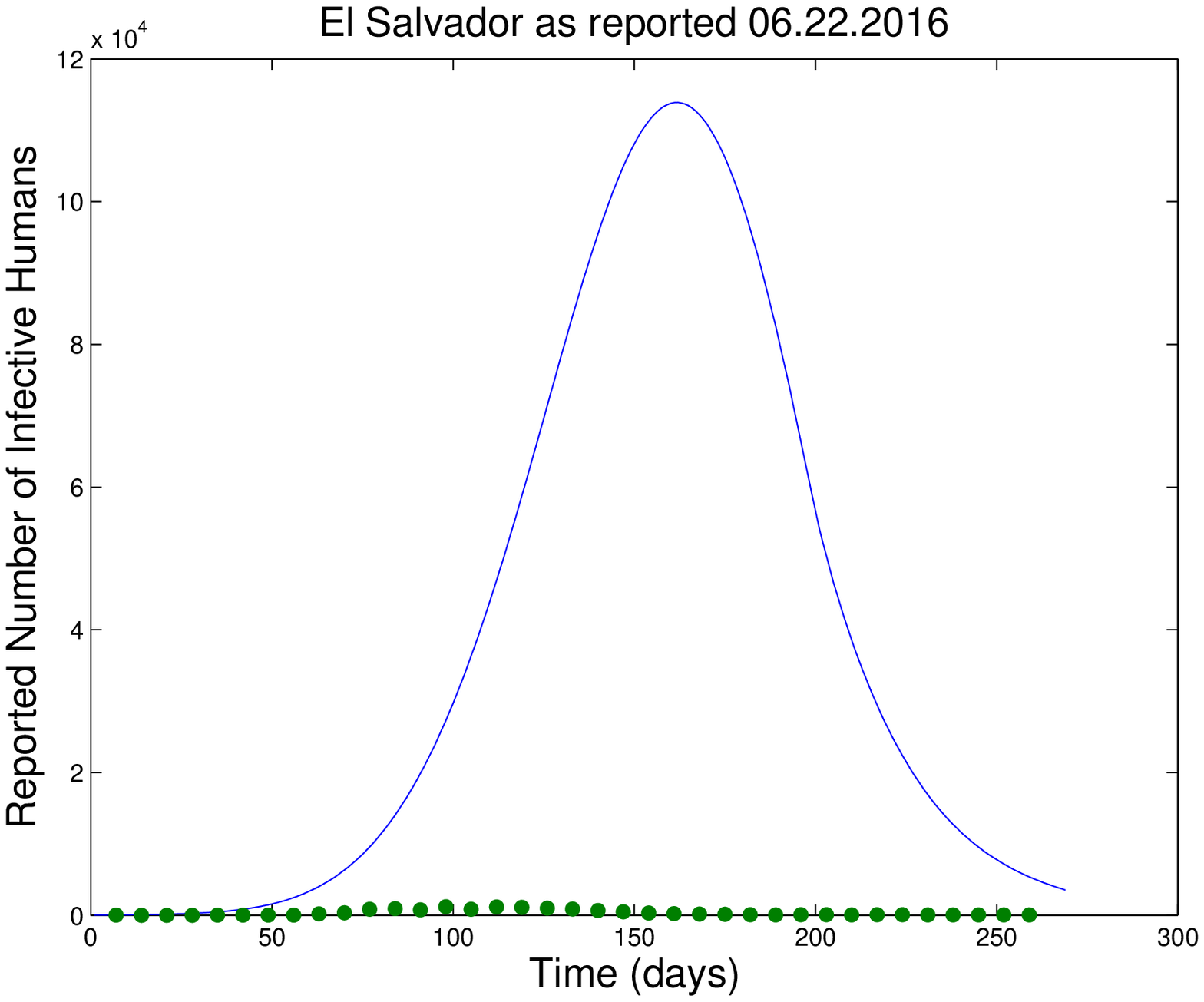}}
\subfigure[Suriname]{\includegraphics[width=.35\textwidth]{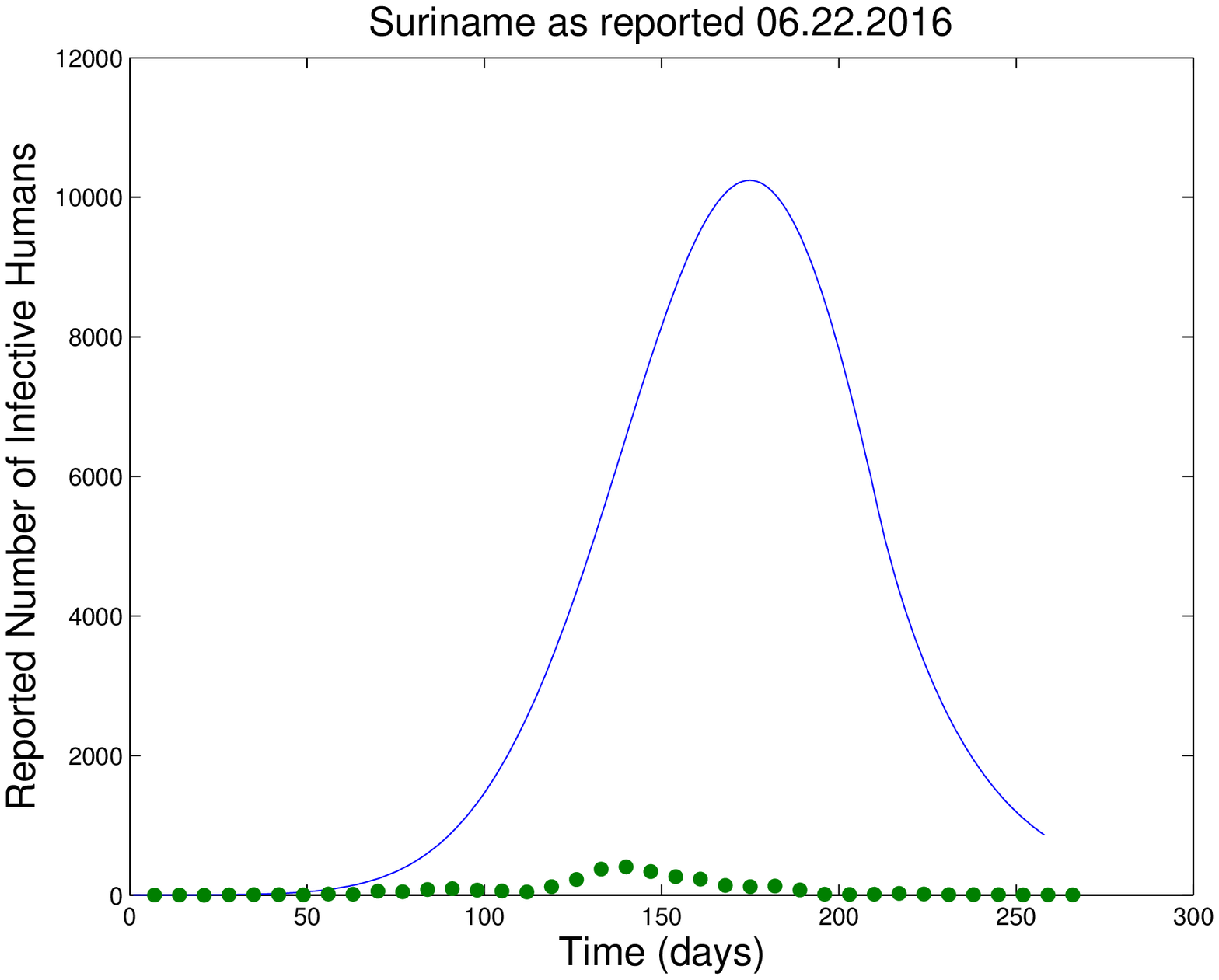}}
\caption{The time series for the number of reported infectives as calculated using Model (1) with the initial number of susceptible people equal to the total population (solid blue line) as compared to the PAHO data (dotted green) as reported 06.22.2016 per country.  The total populations as reported for 2015 are 48,010,049 (Colombia),  6,117,145 (El Salvador), 540,612 (Suriname) accessed 6.22.16 at http://countrymeters.info/.}
\label{wayoff}
\end{figure}

\begin{figure}[H]
\centering
\includegraphics[height=.5\textwidth]{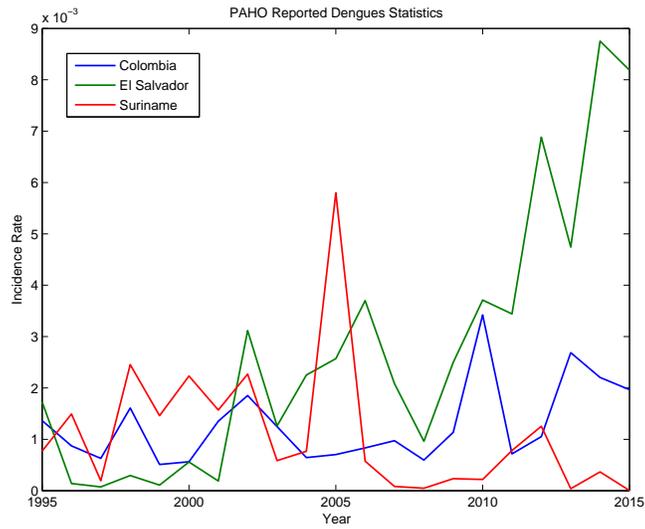}
\caption{Reported rates of incidence are calculated based on the number of reported cases documented \cite{PAHO2} divided by the reported country population in the corresponding year \cite{countrypop}.}
\label{incidencegraph}
\end{figure}

\subsection{Other parameter distributions}

\begin{figure}[H]
\centering
\subfigure[El Salvador - $1/\sigma_v$]{\includegraphics[width=.3\textwidth]{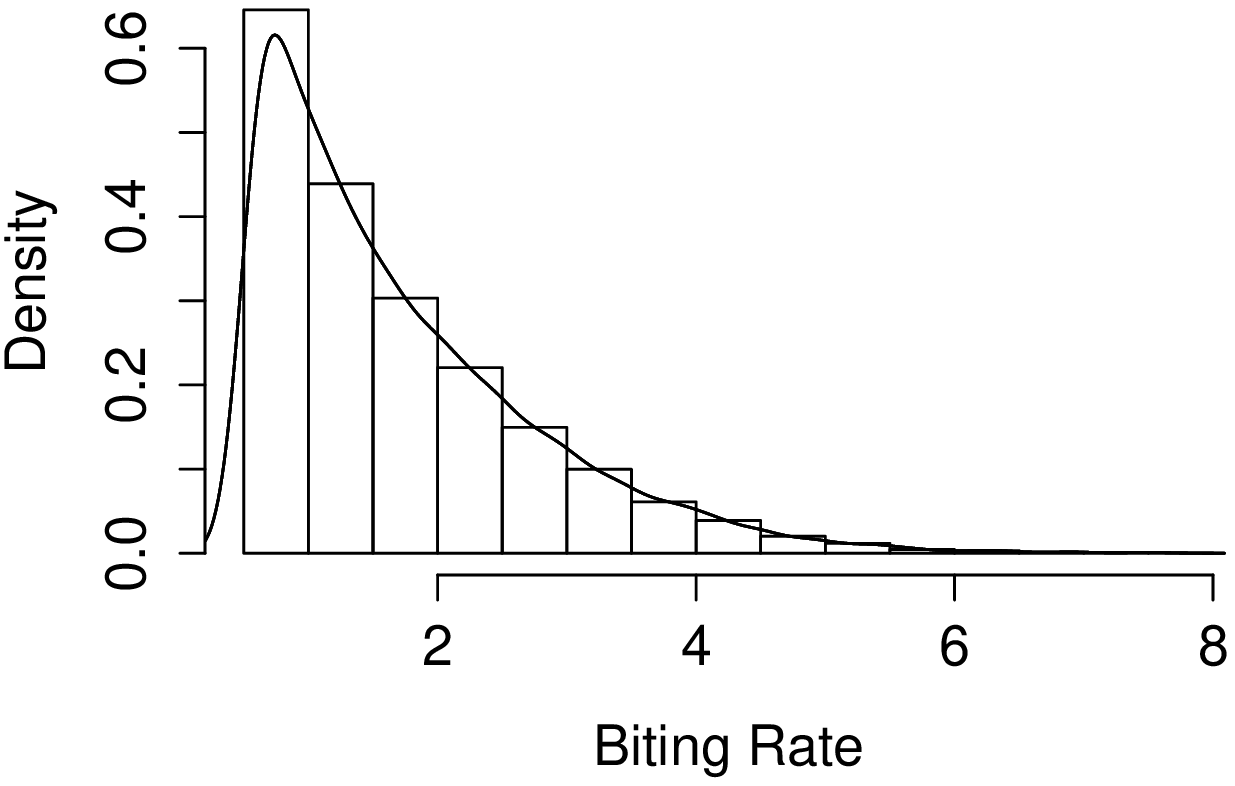}}
\subfigure[El Salvador - $\beta_{hv}$]{\includegraphics[width=.3\textwidth]{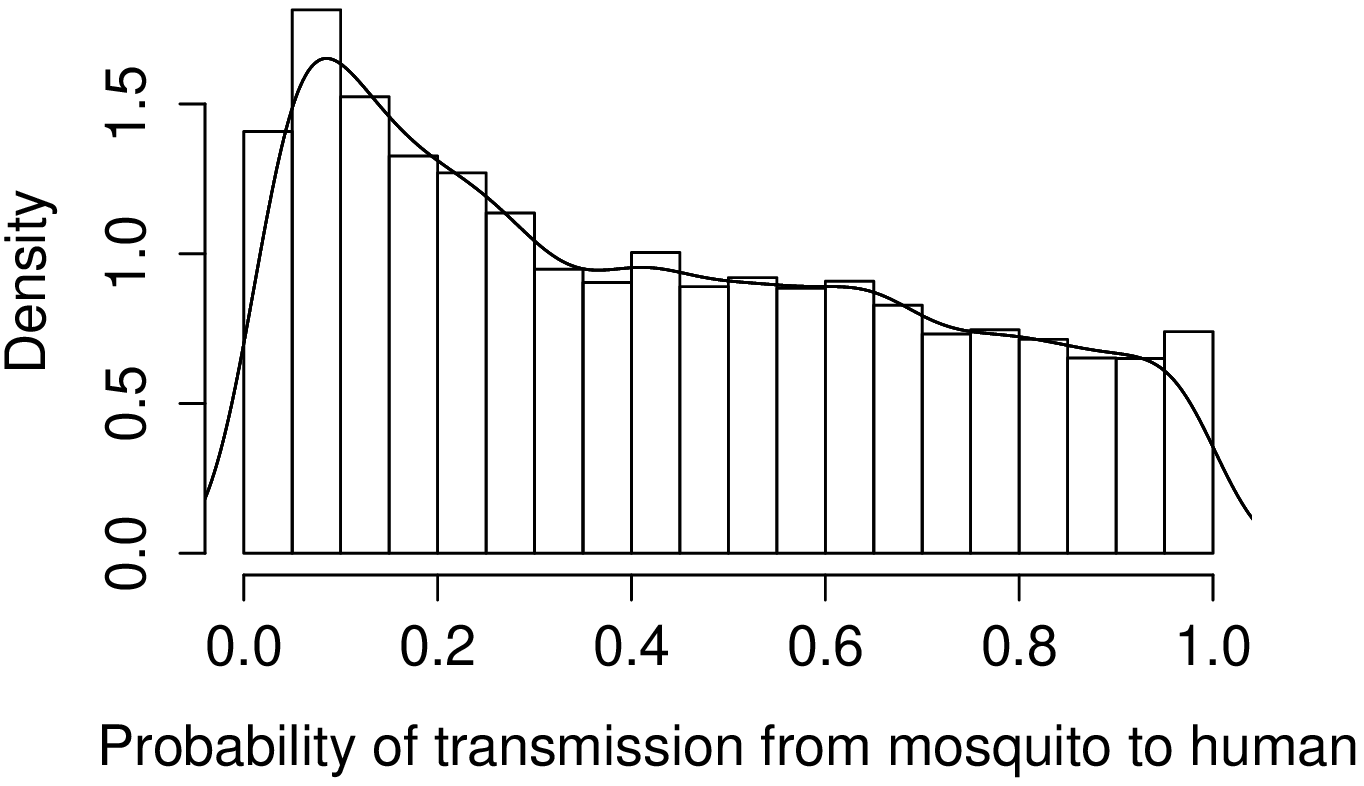}}
\subfigure[El Salvador - $1/\nu_h$]{\includegraphics[width=.3\textwidth]{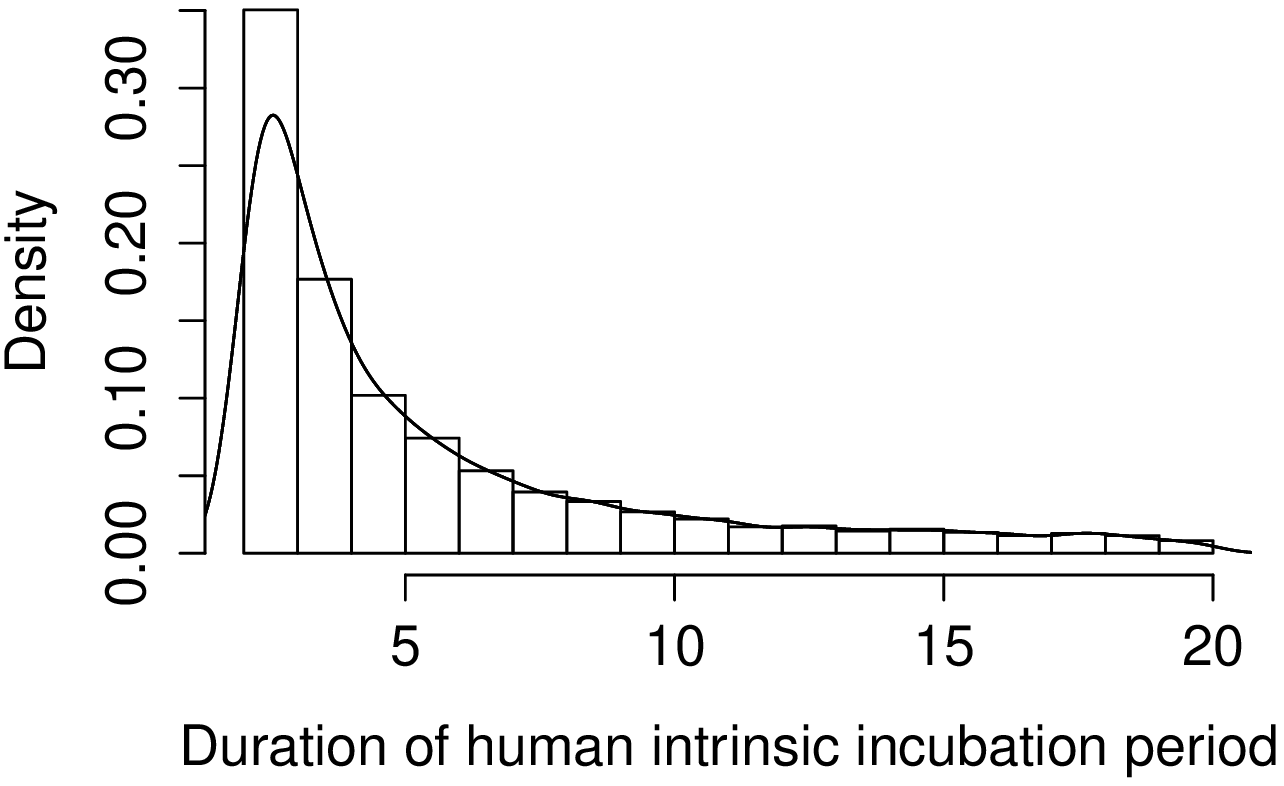}}
\subfigure[El Salvador - $\beta_{vh}$]{\includegraphics[width=.3\textwidth]{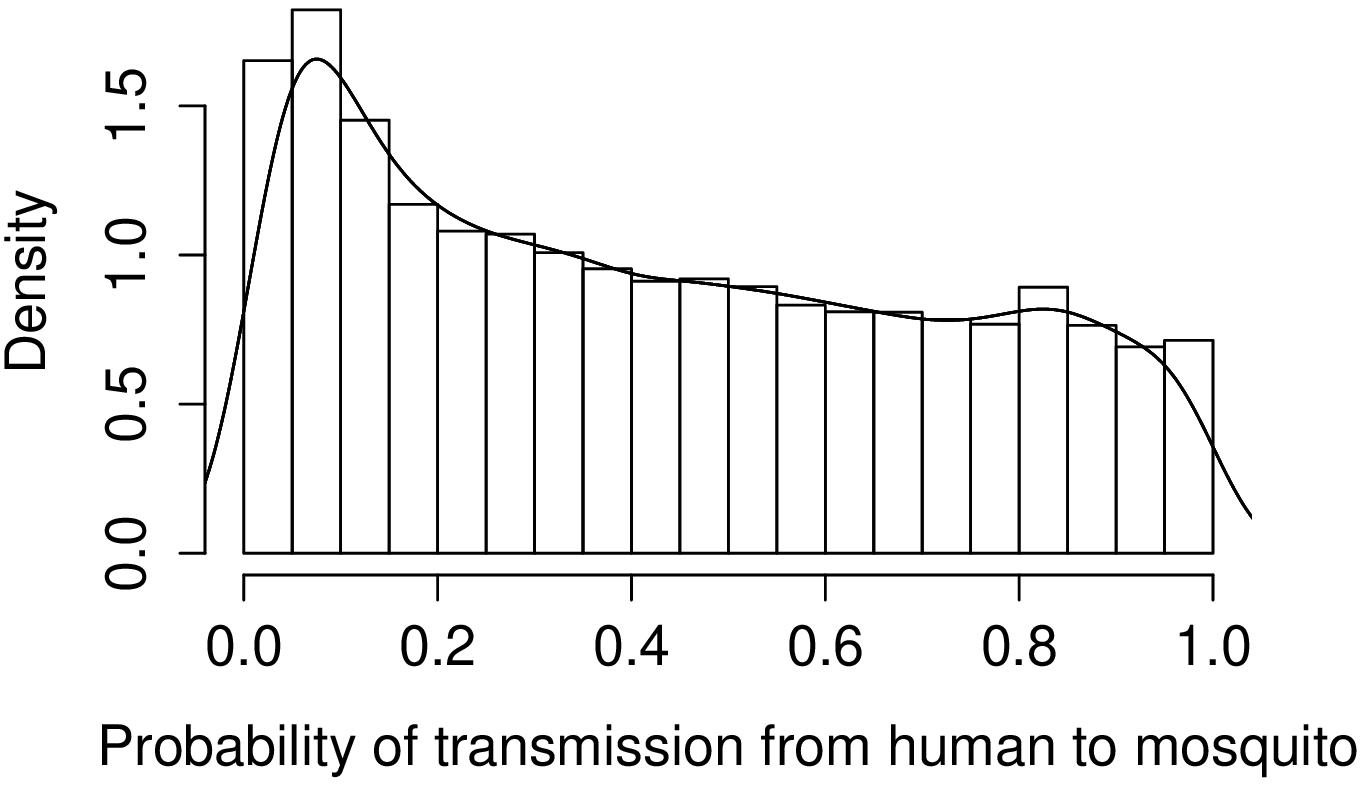}}
\subfigure[El Salvador - $1/\nu_v$]{\includegraphics[width=.3\textwidth]{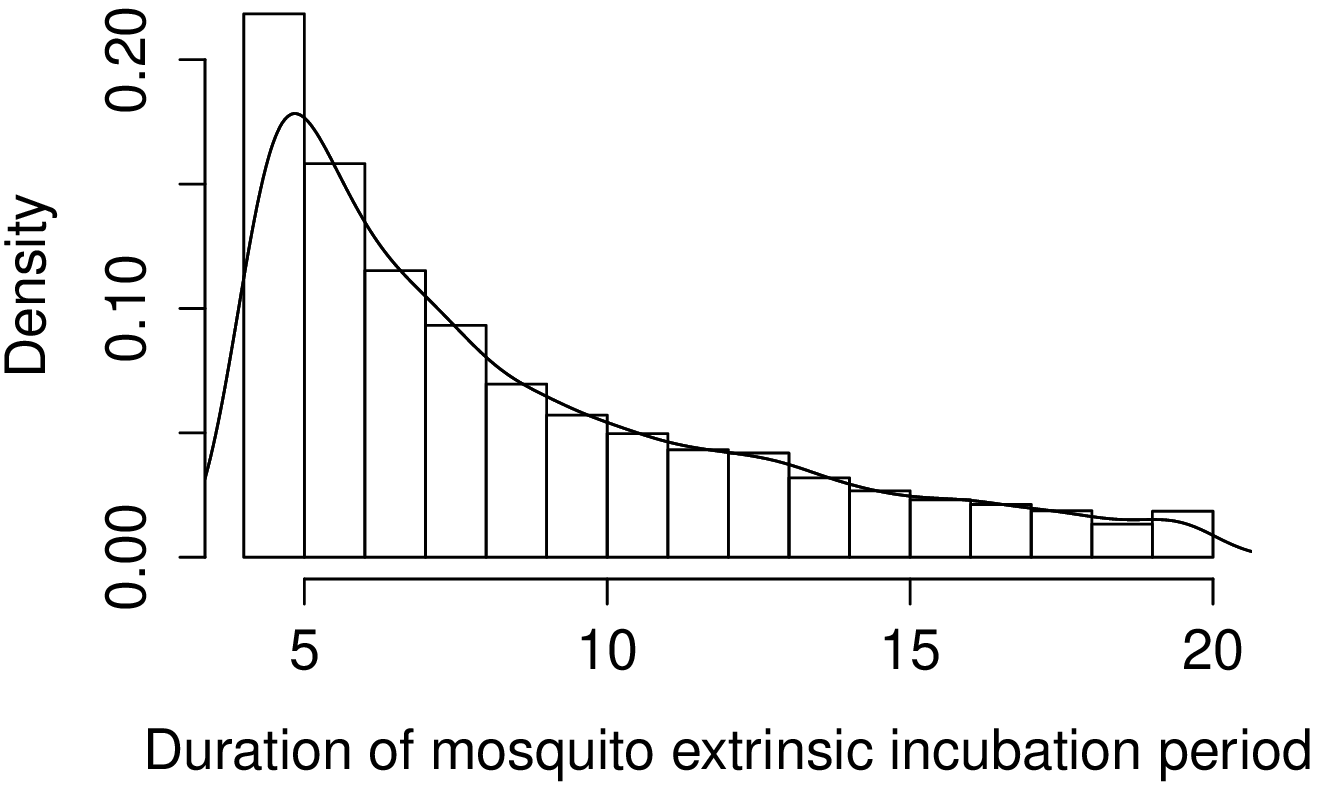}}
\subfigure[El Salvador - $1/\mu_v$]{\includegraphics[width=.3\textwidth]{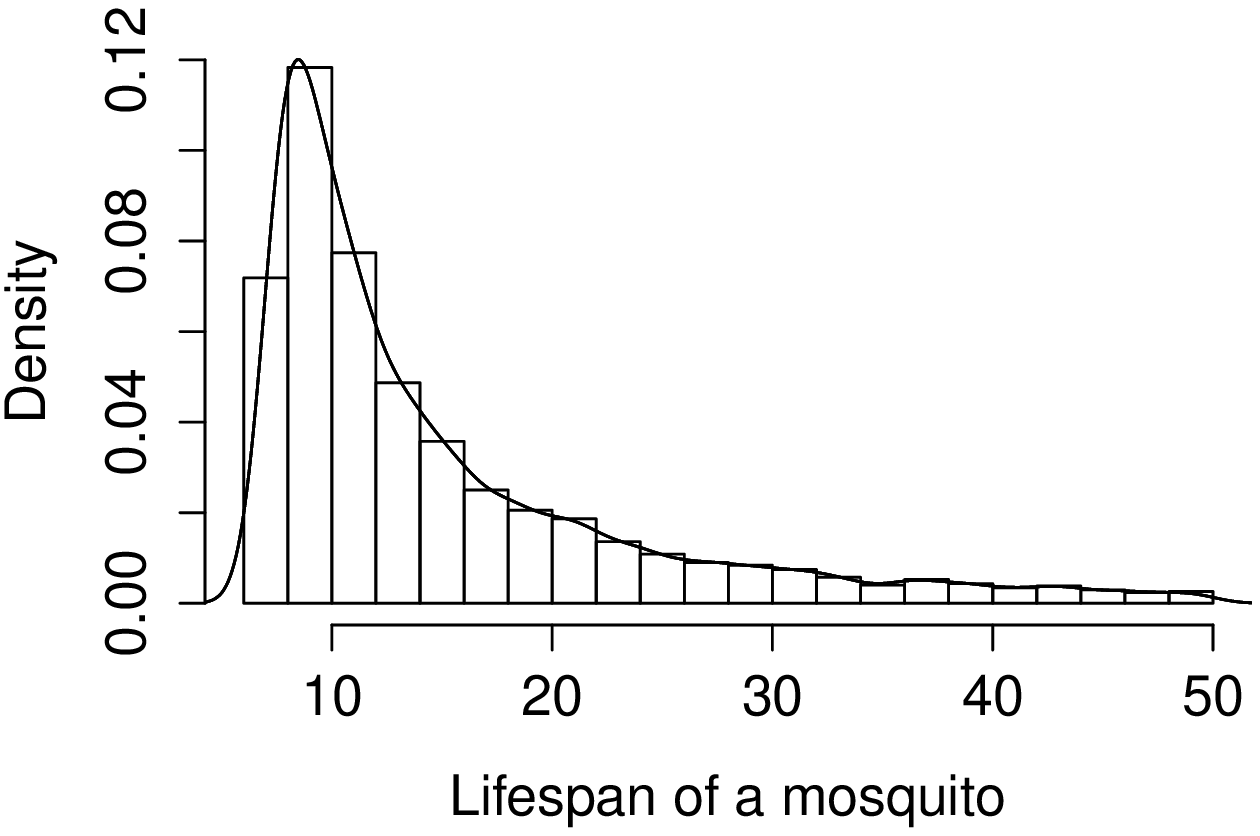}}
\subfigure[El Salvador - $\frac{\nu_v}{\mu_v+\nu_v}$]{\includegraphics[width=.3\textwidth]{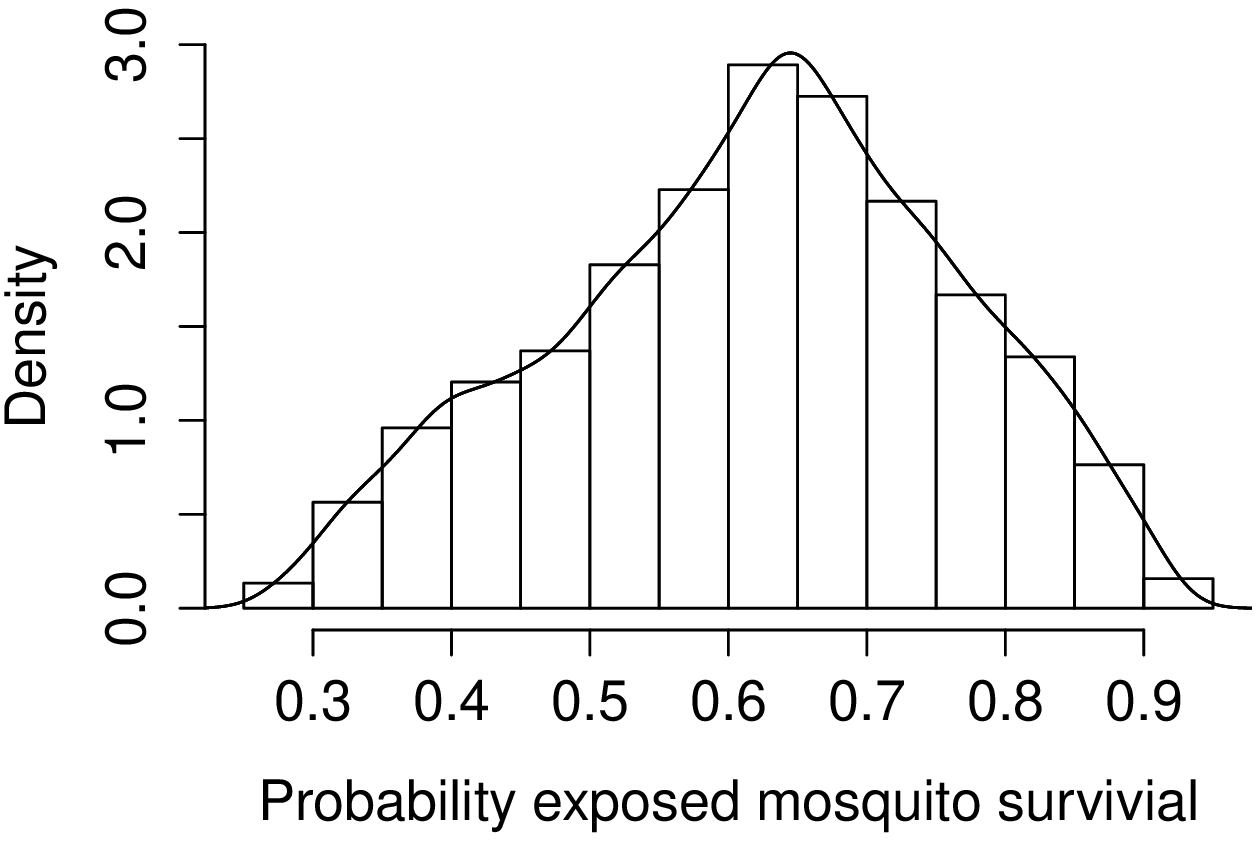}}
\caption{Distributions for the various parameters used for modeling Zika dynamics in El Salvador}
\end{figure}

\begin{figure}[H]
\centering
\subfigure[Suriname - $1/\sigma_v$]{\includegraphics[width=.3\textwidth]{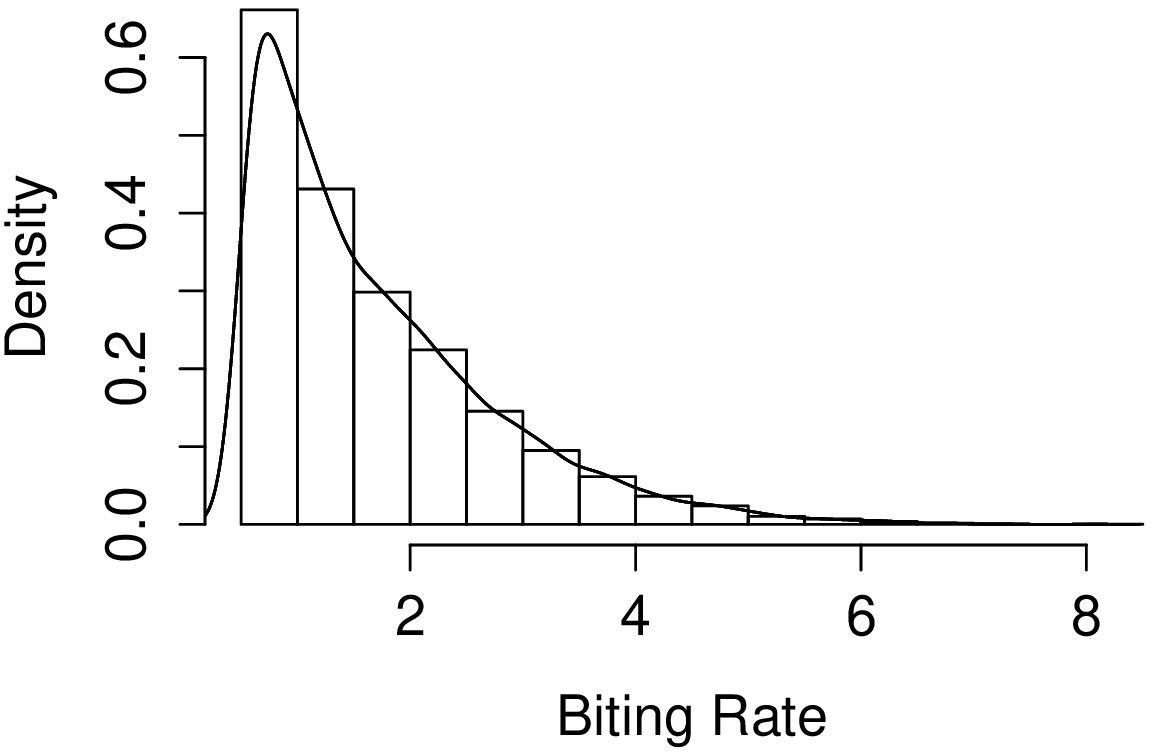}}
\subfigure[Suriname - $\beta_{hv}$]{\includegraphics[width=.3\textwidth]{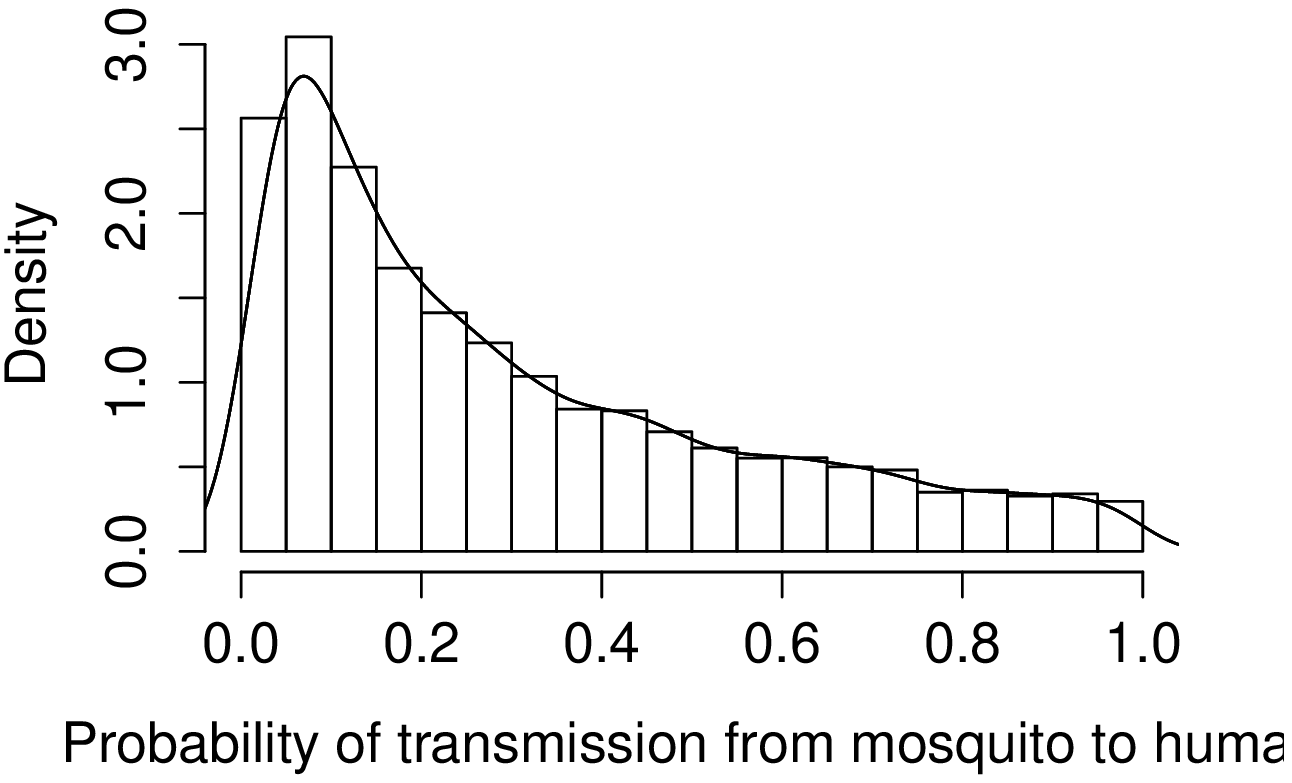}}
\subfigure[Suriname - $1/\nu_h$]{\includegraphics[width=.3\textwidth]{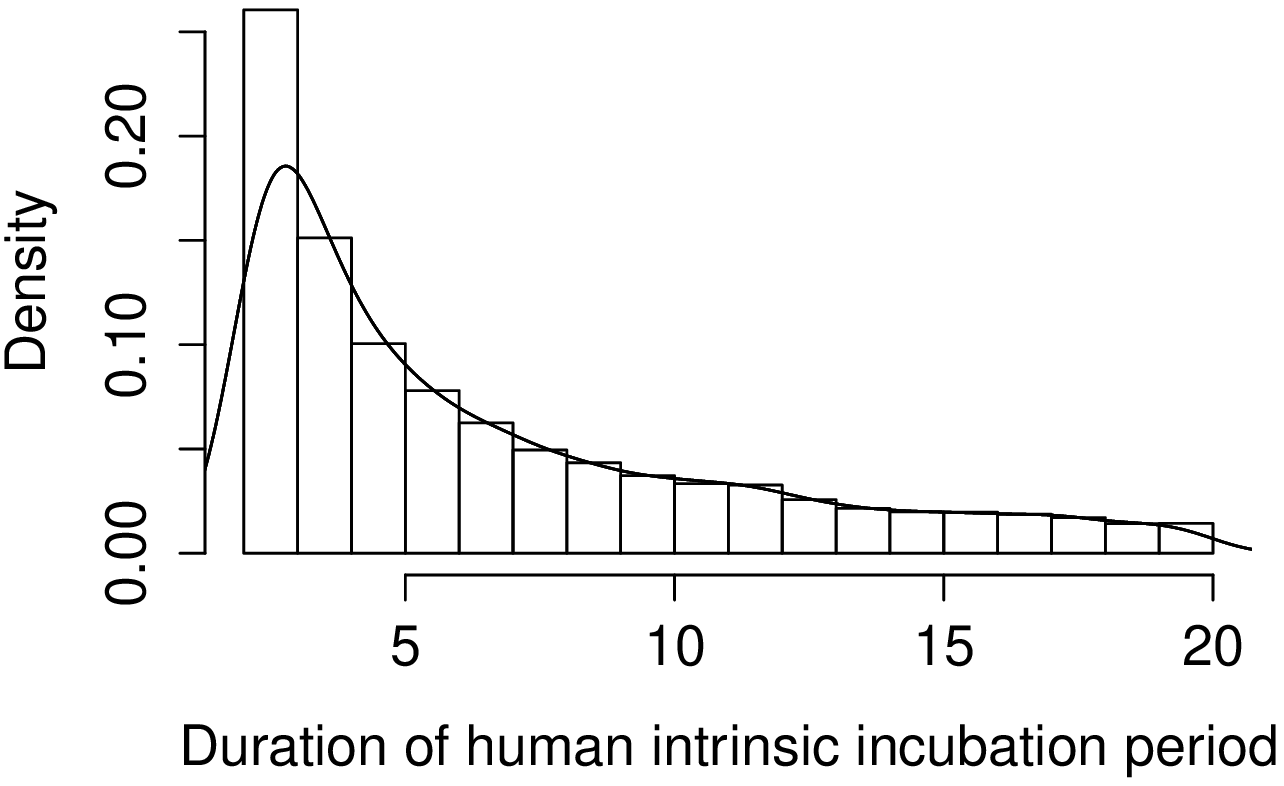}}
\subfigure[Suriname - $\beta_{vh}$]{\includegraphics[width=.3\textwidth]{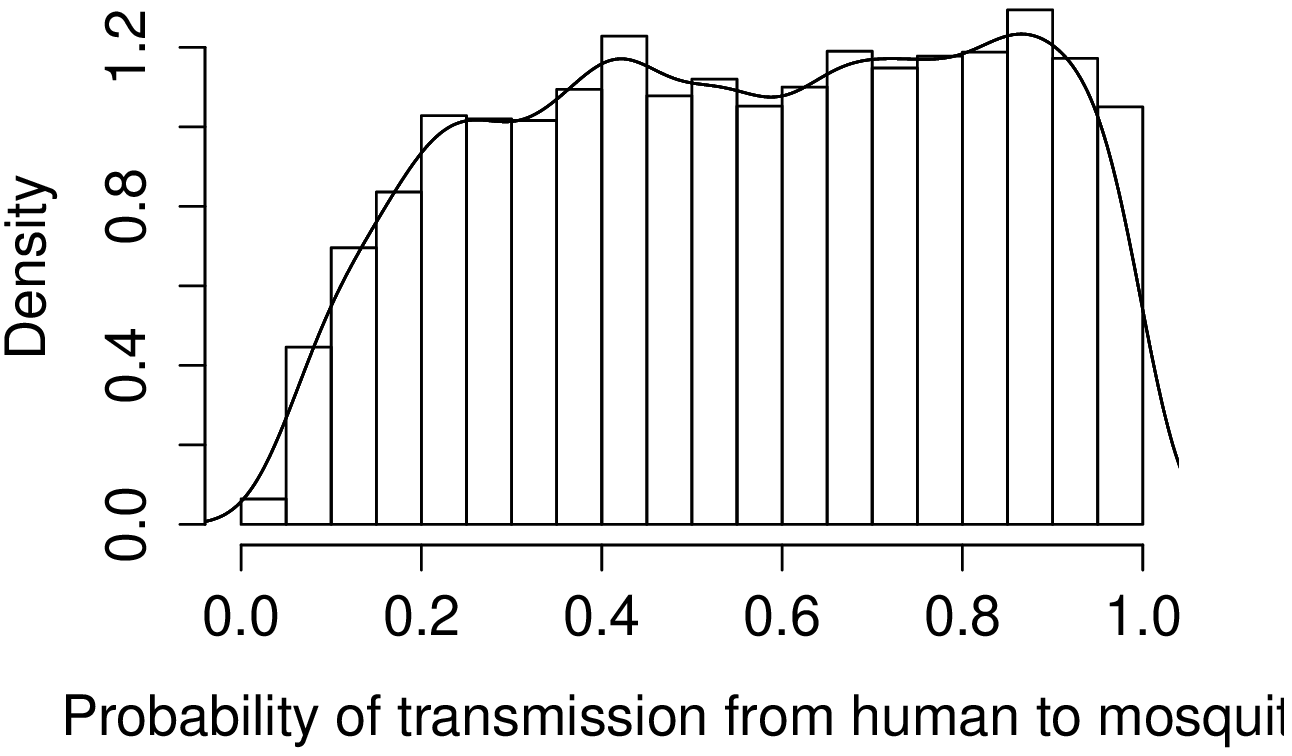}}
\subfigure[Suriname - $1/\nu_v$]{\includegraphics[width=.3\textwidth]{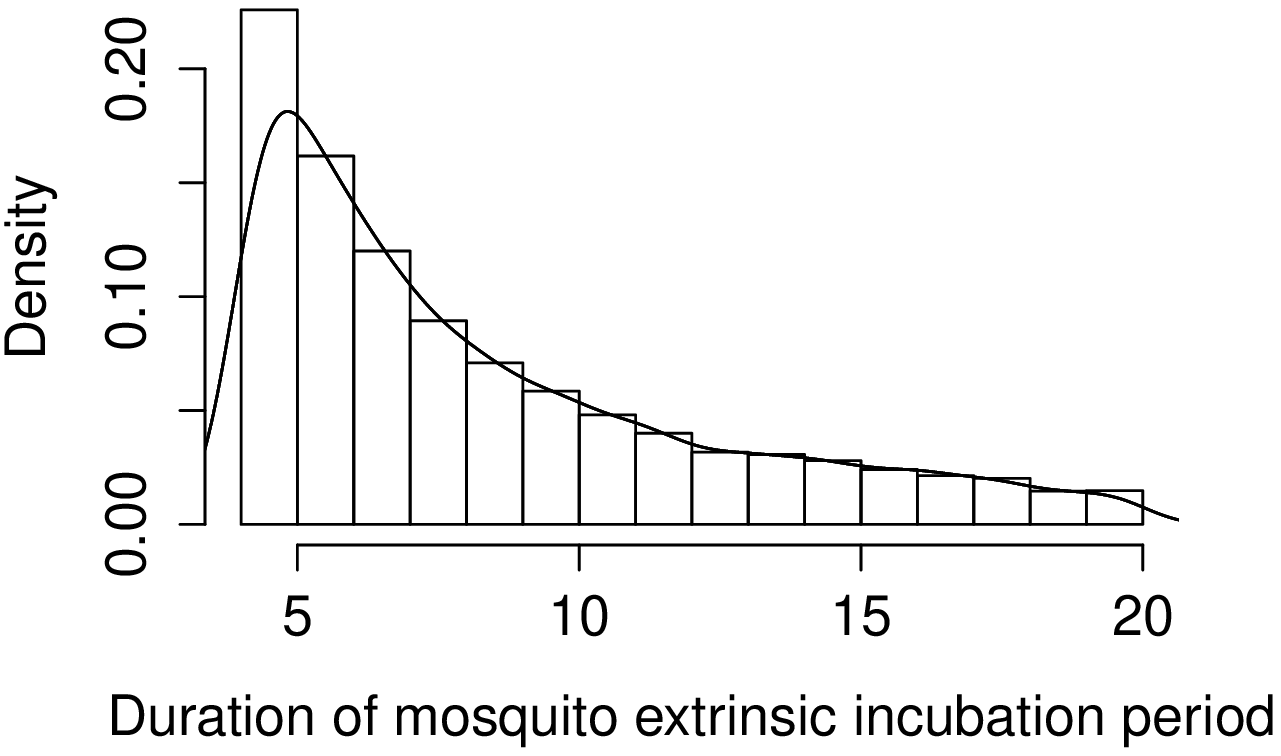}}
\subfigure[Suriname - $1/\mu_v$]{\includegraphics[width=.3\textwidth]{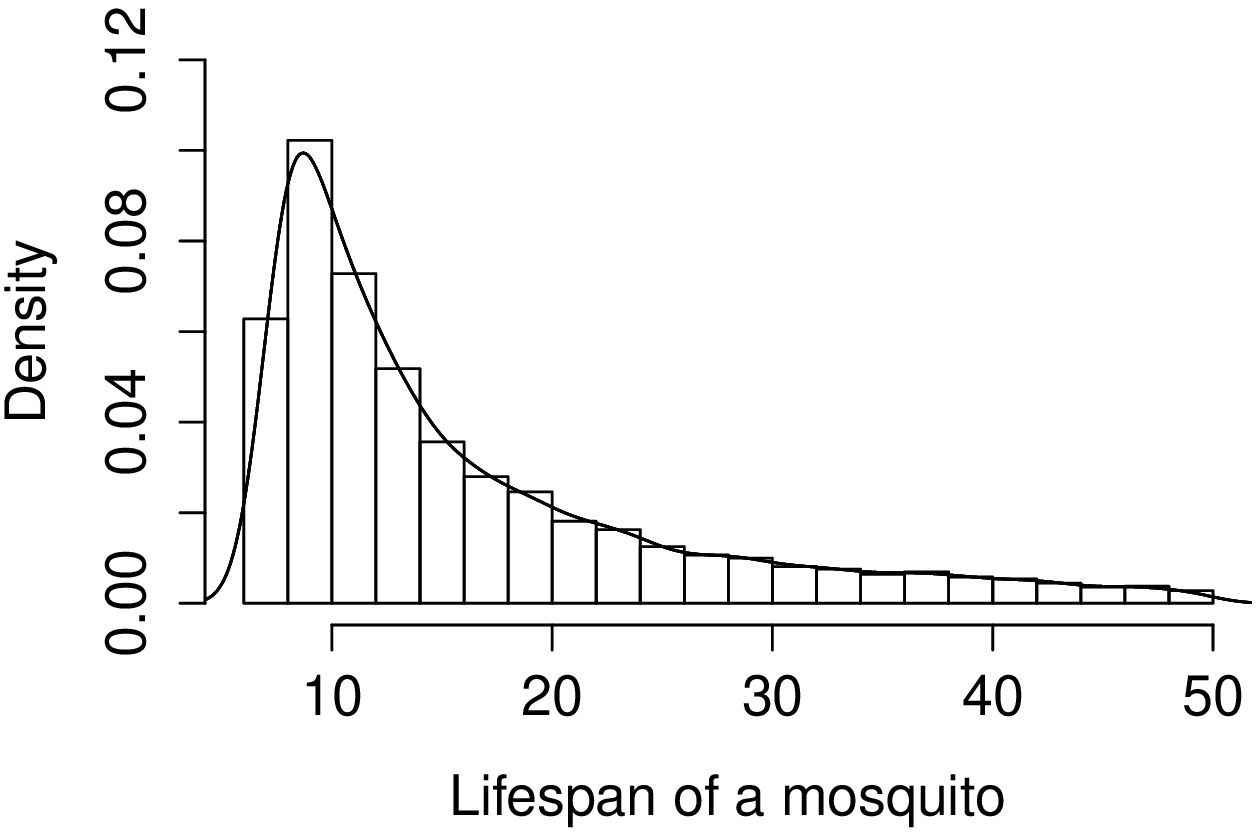}}
\subfigure[Suriname - $\frac{\nu_v}{\mu_v+\nu_v}$]{\includegraphics[width=.35\textwidth]{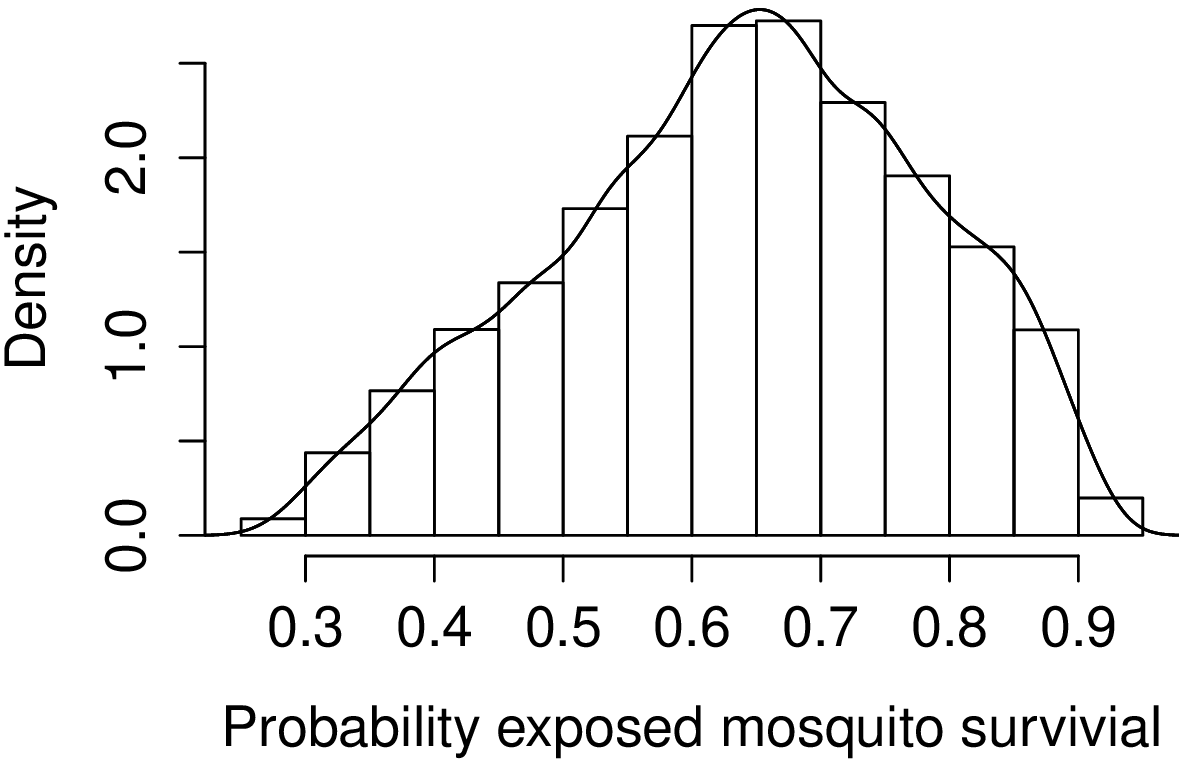}}
\caption{Distributions for the various parameters used for modeling Zika dynamics in Suriname}
\end{figure}

\afterpage{%
    \clearpage
    \begin{landscape}
        \centering
\
\vspace{0.5in}

\begin{table}[H]
\caption{Statistics for the Other Parameter Values for El Salvador}
\label{otherstats_elsal}
\vspace{0.1in}
\begin{footnotesize}
\begin{tabular}{cccccccc}
\hline
& $1/\sigma_v$ & $\beta_{hv}$ & $1/\nu_h$  & $\beta_{vh}$ & $1/\nu_v$ & $1/\mu_v$ &$\frac{\nu_v}{\mu_v+\nu_v}$\\
\hline
\multicolumn{7}{l}{{\color{white}Dependent variables / Populations}} \\
mean &  $1.6894$ &  $0.4201$ &  $5.5626$&  $0.4260$&  $8.4171$& $14.9655$ &  $0.6206$\\
median &  $1.3874$ &  $0.3817$ &  $3.8095$&  $0.3882$&  $7.0799$& $11.4532$ &  $0.6315$ \\
mode  &  $0.7397$ &  $0.0851$ &  $2.5506$&  $0.0760$&  $4.8253$& $8.4276$ &  $0.6451$ \\
$95\%$ C.I. & [0.5004, 3.8294] & [0.01274, 0.9302] & [2.0001, 15.5402] & [0.0104, 0.9289] & [4.0001, 17.0355] & [7.1432, 35.9220] & [0.3475, 0.8832]\\
\\
\hline
\end{tabular}
\end{footnotesize}
\end{table}

\vspace{0.5in}

\begin{table}[H]
\caption{Statistics for the Other Parameter Values for Suriname}
\label{otherstats_elsal}
\vspace{0.1in}
\begin{footnotesize}
\begin{tabular}{cccccccc}
\hline
& $1/\sigma_v$ & $\beta_{hv}$ & $1/\nu_h$  & $\beta_{vh}$ & $1/\nu_v$ & $1/\mu_v$ &$\frac{\nu_v}{\mu_v+\nu_v}$\\
\hline
\multicolumn{7}{l}{{\color{white}Dependent variables / Populations}} \\
mean &  $1.6819$ &  $0.3004$ &  $6.6144$&  $0.5602$&  $8.3111$& $16.1174$ &  $0.6369$\\
median &  $1.3684$ &  $0.2141$ &  $4.8592$&  $0.5657$&  $6.9234$& $12.4293$ &  $0.6453$ \\
mode  &  $0.7375$ &  $0.0702$ &  $2.7695$&  $0.8657$&  $4.8375$& $8.7023$ &  $0.6521$ \\
$95\%$ C.I. & [0.5002, 3.8328] & [0.0101, 0.8456] & [2.0001, 16.8145] & [0.1358, 0.9985] & [4.0000, 16.9617] & [7.1467, 38.2319] & [0.3699, 0.8961]\\
\\
\hline
\end{tabular}
\end{footnotesize}
\end{table}

\end{landscape}
    \clearpage
}

\subsection{Other Reproductive Number distributions}
\begin{figure}[H]
\centering
\subfigure[El Salvador - $R_{hv}$]{\includegraphics[width=.35\textwidth]{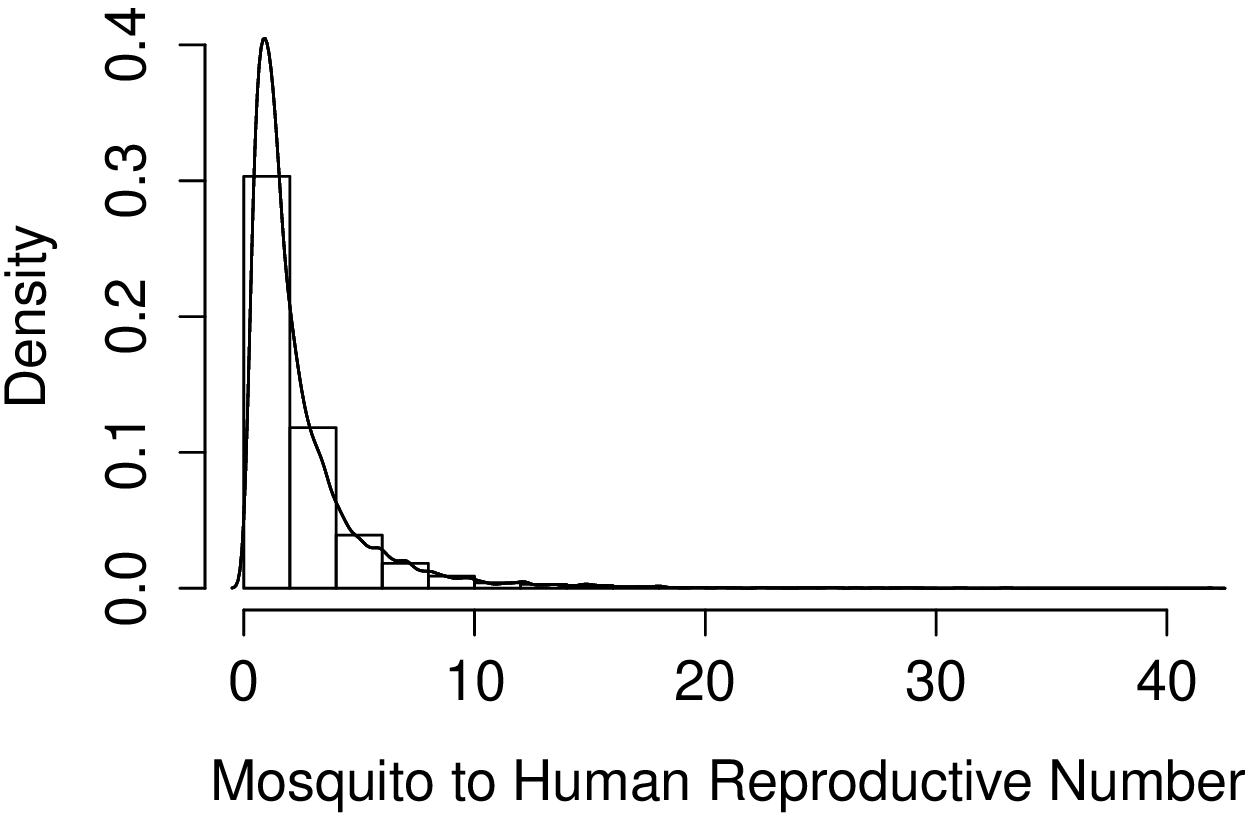}}
\subfigure[El Salvador - $R_{vh}$]{\includegraphics[width=.35\textwidth]{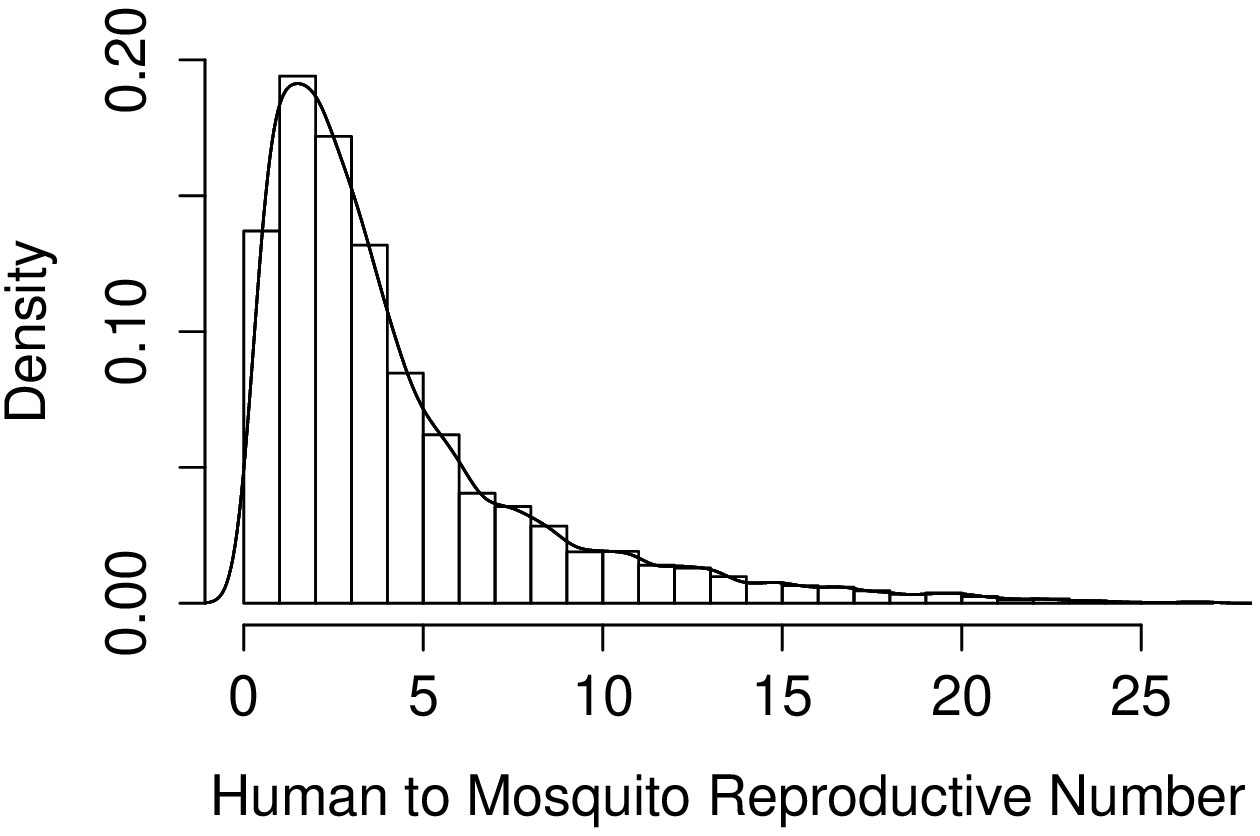}}
\subfigure[Suriname - $R_{hv}$]{\includegraphics[width=.35\textwidth]{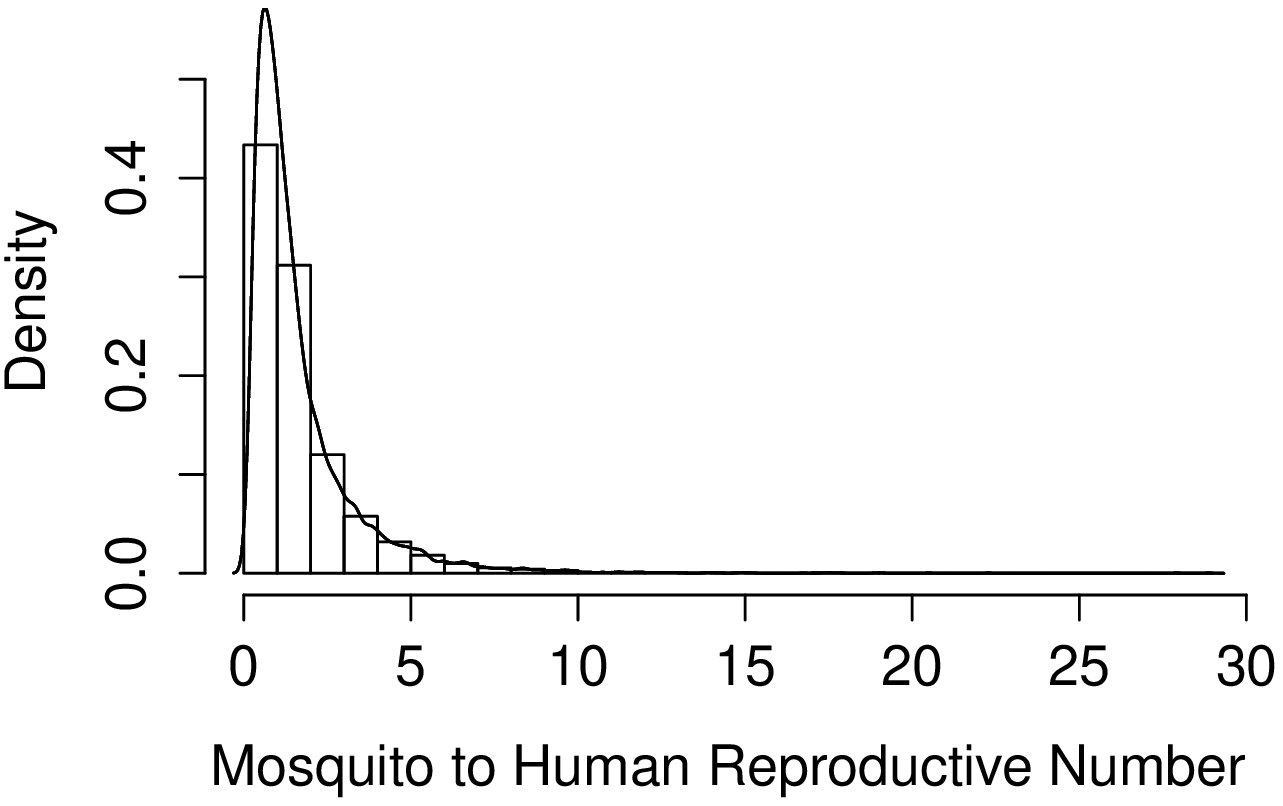}}
\subfigure[Suriname - $R_{vh}$]{\includegraphics[width=.35\textwidth]{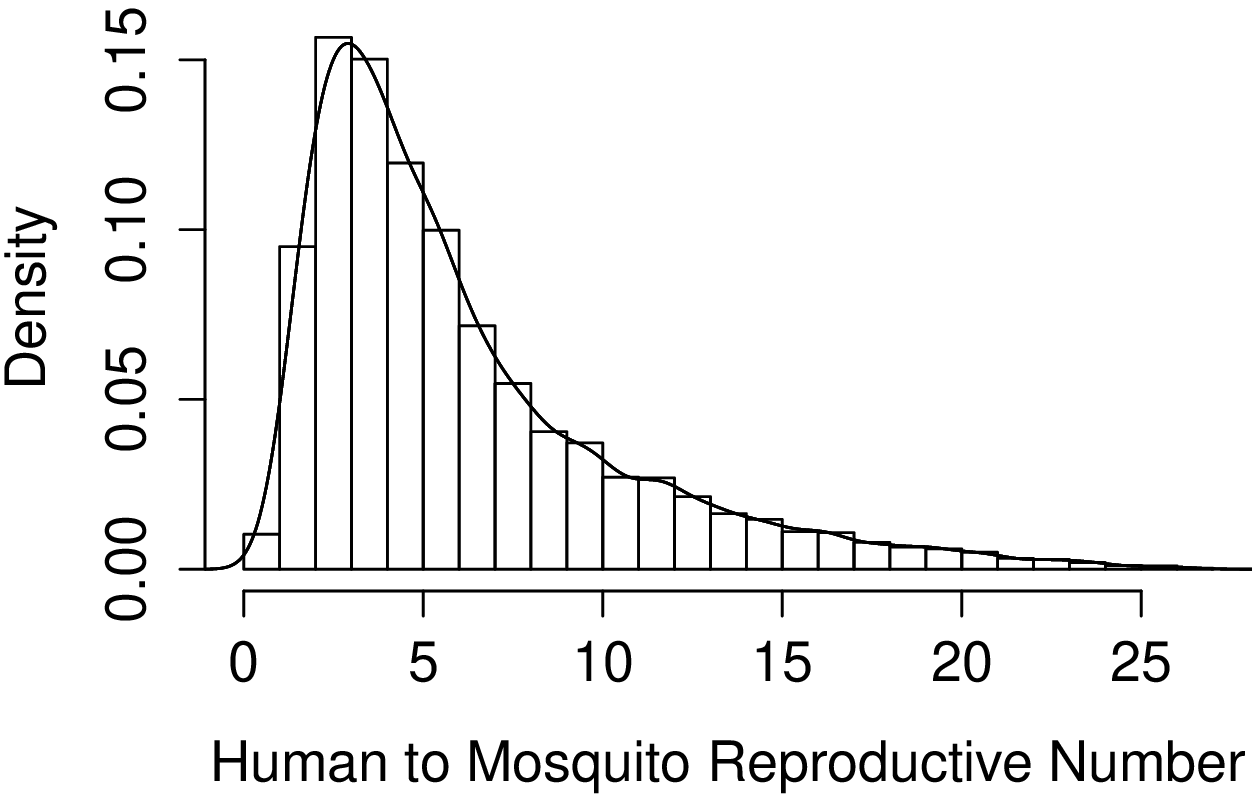}}
\caption{Displayed are the individual to individual reproductive ratios.  For El Salvador the mosquito to human ratio histogram and kernel density is (a) while the human to mosquito ratio histogram and kernel density is (b).  (c) and (d) show the histogram and kernel density plots of the mosquito to human and human to mosquito ratios for Suriname, respectively.}
\end{figure}

\begin{table}[H]
\centering
\caption{Statistics for the Individual to Individual Reproductive Numbers}
\label{otherR0}
\vspace{0.1in}
\begin{footnotesize}
\begin{tabular}{p{2cm}p{3cm}p{3cm}p{3cm}p{3cm}}
\hline
\quad~~  & El Salvador - $R_{hv}$ & El Salvador - $R_{vh}$ & Suriname - $R_{hv}$ & Suriname - $R_{vh}$\\
\hline
\multicolumn{3}{l}{{\color{white}Dependent variables / Populations}} \\
mean & \quad $2.4588$ & \quad $4.2743$ & \quad $1.6667$ & \quad $6.0925$ \\
median & \quad $1.5509$ & \quad $2.9813$ & \quad $1.1414$ & \quad $4.7115$ \\
mode &\quad $0.9238$ & \quad $1.5273$ & \quad $0.6250$ & \quad $2.8961$\\
$95\%$ C.I. & [0.1519, 7.4016] & [0.1663, 12.9415] & [0.1473, 4.8498] &  [0.6002, 15.7379]\\
\\
\hline
\end{tabular}
\end{footnotesize}
\end{table}

\bibliographystyle{acm}
\bibliography{zikamain}

\end{document}